\documentclass[11pt,aaspp4]{aastex}

\slugcomment{\it Accepted to the Astronomical Journal}
\lefthead{Griffith \& Stern}
\righthead{AGN Morphologies}

\usepackage{graphicx} % needed for including graphics e.g. EPS, PS
\usepackage{amssymb}
\usepackage{amsmath}
\usepackage{anysize}
\usepackage{rotating}
\usepackage{multirow}
\usepackage{float}

\def\spose#1{\hbox to 0pt{#1\hss}}
\def\simlt{\mathrel{\spose{\lower 3pt\hbox{$\mathchar"218$}}
     \raise 2.0pt\hbox{$\mathchar"13C$}}}
\def\simgt{\mathrel{\spose{\lower 3pt\hbox{$\mathchar"218$}}
     \raise 2.0pt\hbox{$\mathchar"13E$}}}

\begin{document}

\title{Morphologies of Radio, X-Ray, and Mid-Infrared Selected AGN\altaffilmark{*}}

\author{R\textsc{oger} L. G\textsc{riffith}\altaffilmark{1} \&
D\textsc{aniel} S\textsc{tern}\altaffilmark{1}}

\altaffiltext{*}{Based on observations with: the NASA/ESA {\it
Hubble Space Telescope}, obtained at the Space Telescope Science
Institute, which is operated by AURA Inc, under NASA contract NAS
5-26555; Subaru Telescope, which is operated by the National
Astronomical Observatory of Japan; {\it XMM-Newton}, an ESA science
mission with instruments and contributions directly funded by ESA
Member States and NASA under large program 175.A-0839; National
Radio Astronomy Observatory, which is a facility of the National
Science Foundation operated under cooperative agreement by Associated
Universities; and the {\it Spitzer Space Telescope}, which is
operated by the Jet Propulsion Laboratory, California Institute of
Technology, under NASA contract 1407.}

\altaffiltext{1}{Jet Propulsion Laboratory, California Institute
of Technology, MS 169-327, 4800 Oak Grove Dr., Pasadena, CA 91109}


\begin{abstract}

We investigate the optical morphologies of candidate active galaxies
identified at radio, X-ray, and mid-infrared wavelengths.  We use
the Advanced Camera for Surveys General Catalog (ACS-GC) to identify
372, 1360, and 1238 AGN host galaxies from the VLA, {\it XMM-Newton}
and {\it Spitzer Space Telescope} observations of the COSMOS field,
respectively.  We investigate both quantitative ({\tt GALFIT}) and
qualitative (visual) morphologies of these AGN host galaxies, split
by brightness in their selection band.  We find that the samples
are largely distinct, though extensive overlap exists between certain
samples, most particularly for the X-ray and mid-IR selected sources
with unresolved optical morphologies.  We find that the radio-selected
AGN are most distinct, with a very low incidence of having unresolved
optical morphologies and a high incidence of being hosted by
early-type galaxies.  In comparison to X-ray selected AGN, mid-IR
selected AGN have a slightly higher incidence of being hosted by
disk galaxies.  These morphological results conform with the results
of \citet{hic09} who studied the colors and large-scale clustering
of AGN, and found a general association of radio-selected AGN with
``red sequence'' galaxies, mid-IR selected AGN with ``blue cloud''
galaxies, and X-ray selected AGN straddling these samples in the
``green valley.''  We also find that optical brightness scales with
X-ray and mid-IR brightness, while little correlation is evident
between optical and radio brightness.  This suggests that X-ray and
mid-IR selected AGN have similar Eddington ratios, while radio-selected
AGN represent a different accretion mechanism with a lower and wider
range of Eddington ratios.  In the general scenario where AGN
activity marks and regulates the transition from late-type disk
galaxies into massive elliptical galaxies, this work suggests that
the earlier stages are most evident as mid-IR selected AGNs.  Mid-IR
emission is less susceptible to absorption than the relatively soft
X-rays probed by {\it XMM-Newton}, which are seen at later stages
in the transition.  Radio-selected AGN are then typically associated
with minor bursts of activity in the most massive galaxies.

\end{abstract}

\section{Introduction}

In order to derive a coherent view of galaxy evolution we must
investigate the different modes by which galaxies evolve. One of
these modes is AGN activity in galaxies. There appears to be a
strong connection between galaxies and their super massive black
holes (SMBHs) as evidenced in the local universe by observed
correlations between central SMBH masses and host galaxy properties,
including luminosity (\citealt{kor95}; \citealt{mar03}), velocity
dispersion (\citealt{fer00}; \citealt{geb00}), and mass (\citealt{mag98}).
These observations imply that galaxies and their SMBHs co-evolve
in a systematic, yet not well understood process.  Studying the
morphologies of AGN host galaxies can provide valuable clues into
the formation and evolution of these systems and may provide a
general understanding of galaxy evolution.

At low redshifts ($0.03 \le z \le 0.3$), \cite{kau03} explore the
properties of a large sample of narrow-lined AGN host galaxies from
the Sloan Digital Sky Survey \citep[SDSS;][]{yor00} and find that
AGN host galaxies have similar properties to ordinary early-type
galaxies, albeit with younger stellar populations indicating recent
episodes of star formation.  Studying SDSS images of $\sim 100$ of
the brightest AGN at $z < 0.1$, \cite{kau03} find that they are
hosted by approximate equal numbers of disk, spheroid, and interacting
galaxies.

At higher redshift, closer to the peak of AGN activity, morphological
studies of AGN host galaxies require the spatial resolution of the
{\it Hubble Space Telescope}.  To date, most large studies have
only considered X-ray selected AGN.  \cite{gab09}, studying a sample
of $\sim 400$ spectroscopically-confirmed AGN at $0.3 \le z \le
1.0$ in the Cosmic Evolution Survey (COSMOS) \citep{scoville07}
field, conclude that X-ray AGN span a substantial range of morphologies
which peak between bulge-dominated, early-type galaxies and
disk-dominated, late-type galaxies.  They suggest the AGN activity
coincides with the transition between disk-dominated to bulge-dominated
galaxies, and might not be triggered by major mergers.  \cite{geo09}
performed visual classifications on a sample of 454 X-ray selected
AGN in the Great Observatories Origins Deep Survey (GOODS) \citep{gia04}
and All-wavelength Extended Groth strip International Survey (AEGIS)
fields.  Their sample was restricted to the redshift range $0.5 \le
z \le 1.3$, and they find that disk-dominated hosts contribute $30
\pm 9$\%\ to the total AGN space density and $23 \pm 6$\%\ to the
luminosity density since $z \sim 1$.  They conclude that AGN in
disk galaxies are likely fueled by minor interactions and disk
instabilities rather than by a major mergers.

However, previous {\it Hubble} studies have suffered by
typically only considering AGN selected by a single method.  Besides
X-ray selection, AGN are also efficiently identified by radio
luminosity and mid-IR color.  In particular, the relatively soft
($< 10$~keV) photons observed by {\it Chandra} and {\it XMM-Newton}
come from very near the central SMBH and are susceptible to absorption.
Radio and mid-IR emission are less susceptible to absorption, and
thus readily identify both unobscured and obscured AGN.  On the
other hand, only $\sim 10$\% of AGN are radio loud \citep[e.g.,][]{ste00},
and mid-IR selection is only effective for the highest luminosity
AGN \citep[e.g.,][]{donley07, eck10}.  \cite{hic09}, studying the
spatial clustering of 585 AGN at $0.25 < z < 0.8$ in the AGN and
Galaxy Evolution Survey (AGES; Kochanek, in prep.) of the Bo\"otes
field, find that: (a) radio-selected AGN are preferentially found
in luminous red sequence galaxies and are strongly clustered; (b)
X-ray-selected AGN are preferentially found in galaxies which
populate the ``green valley'' of color-magnitude space and are
clustered similarly to normal galaxies; and (c) mid-IR-selected AGN
reside in slightly bluer, less luminous galaxies than the X-ray AGN
and are also weakly clustered.

In this paper, we extend the results of \citet{gab09} and \citet{geo09}
by considering the {\it Hubble} morphologies of AGN selected by
more than their X-ray properties, and we extend the results of
\citet{hic09} by seeing whether the optical morphologies of AGN
selected using different criteria are consistent with the expectations
derived from their clustering properties.  As an input sample, we
consider radio, X-ray, and mid-IR selected AGN identified in the
COSMOS survey.  We use {\it Hubble} data to study the optical
morphologies of the AGN host galaxies both quantitatively (with
{\tt GALFIT}) and qualitatively (visually).  This paper is organized
as follows:  in \S2 we describe the data sets used for this analysis,
followed by a detailed discussion of catalog cross-correlation in
\S3.  Section~4 discusses the methodology of the morphological
classifications and results from this analysis.  During the course
of this investigation, we identified a handful of rare, interesting
AGN, such as the dual AGN published in \citet{com09}, as well as
offset AGN and lensed AGN candidates.  These sources are discussed
in \S5.  Finally, we discuss the results in $\S$6.  Where necessary,
we adopt the concordance $\Lambda$-cosmology with $\Omega_M = 0.3$,
$\Omega_\Lambda = 0.7$, and $H_0 = 70\, {\rm\,km\,s^{-1}\,Mpc^{-1}}$.

\section{The Data}

\subsection{{\it Hubble} ACS imaging}

The cornerstone data set for the COSMOS survey is its wide-field
{\it Hubble} Advanced Camera for Surveys (ACS) imaging \citep{scoville07}.
With 583 single-orbit {\it Hubble} ACS F814W ($I$-band; hereafter
$I814$) observations, it is the largest contiguous {\it Hubble}
imaging survey to date.  The imaging covers an area of approximately
1.8 square degrees in the COSMOS field, centered at RA=10:00:28.6,
DEC=+02:12:21.0 (J2000).  The 50\% completeness limit for sources
0.5\arcsec\ in diameter is $I814_{\rm AB}=26.0$.  The imaging has
a resolution of 0.09\arcsec\ (FWHM) and a pixel scale of 0.05\arcsec\
pix$^{-1}$. The COSMOS survey was designed to study the evolution
of galaxies, AGN, and dark matter in the context of large-scale
structure (LSS). The high resolution imaging allows for the measurement
of galaxy morphological parameters with high accuracy.  A detailed
description of the {\it Hubble} data processing is provided in
\cite{koek07}.

\subsection{VLA imaging}

The VLA-COSMOS large project \citep{schin07} acquired deep, uniform
1.4 GHz data over the entire COSMOS field using the A-array
configuration of the Very Large Array (VLA).  This configuration
provides a resolution of approximately 2\arcsec\ (FWHM) at 1.4 GHz,
which is well-matched to the spatial resolution of the optical and
near-IR ground-based data.  The deep observations provide a mean
rms noise of $\sim 10.5~ \mu$Jy per beam.  A total of 23 separate
pointings were observed to fully cover the COSMOS field. The
VLA-COSMOS project was constructed to study a wide range of
cosmological questions, including the history of star formation and
the growth of super-massive black holes through AGN activity.

\subsection{{\it XMM-Newton} imaging}

The {\it XMM-Newton} COSMOS survey (\citealt{has07}, \citealt{cap09})
acquired deep X-ray data over the entire COSMOS {\it Hubble} ACS
field. A total of 55 pointings were observed (2.13 deg$^2$) for a
total of $\sim$ 1.5 Ms with an average vignetting corrected depth
of 40 ks across the field of view. The survey has flux limits of
$\sim$1.7$\times$10$^{-15}$ erg cm$^{-2}$ s$^{-1}$,
$\sim$9.3$\times$10$^{-15}$ erg cm$^{-2}$ s$^{-1}$ and
$\sim$1.3$\times$10$^{-14}$ erg cm$^{-2}$ s$^{-1}$ over 90\% of the
area (1.92 deg$^2$) in the 0.5-2 keV, 2-10 keV and 5-10 keV energy
bands, respectively. The {\it XMM-Newton} observations are important
for identifying AGN, groups of galaxies, and galaxy clusters. The
primary goal of the {\it XMM-Newton} COSMOS survey was to study the
co-evolution of AGN and galaxies as a function of environment.

\subsection{{\it Spitzer} imaging}

The S-COSMOS survey \citep{san07} is a {\it Spitzer} Legacy program
which carried out a uniformly deep survey of the full COSMOS field
in seven mid-IR bands (3.6, 4.5, 5.8, 8.0, 24, 70, and 160 $\mu$m).
The Infrared Array Camera (IRAC) \citep{faz04} and Multiband Imaging
Photometer for {\it{Spitzer}} (MIPS) \citep{rie04} data are well
suited to addressing the stellar-mass assembly of galaxies (through
IRAC) and  dust-embedded sources such as starburst galaxies and AGN
(through MIPS and IRAC). In particular, heavily-obscured AGNs will
be largely missed from surveys at UV, optical, and soft X-ray
energies. The typical depth of S-COSMOS is between 1.2 and 2.2~ks.
Refer to Table 4 in \cite{san07} for integration times and band
sensitivities.

\subsection{ACS-GC}

The Advanced Camera for Surveys General
Catalog\footnote[1]{ugastro.berkeley.edu/$\sim$~rgriffit/Morphologies/}
(ACS-GC) data (Griffith et al., in prep.) was constructed to study
the evolution of galaxy morphologies over a wide range of look-back
times. The ACS-GC uniformly analyzes the largest {\it Hubble} ACS
imaging surveys (AEGIS, GEMS, GOODS-S, GOODS-N, and COSMOS) using
the {\tt GALAPAGOS} code (Barden et al., in prep.) to measure
quantitative galaxy morphologies for over 490,000 sources. The
COSMOS survey comprises 65.3\% of the catalogued sources. The ACS-GC
data set also combines the large redshift surveys (DEEP2, COMBO17,
zCOSMOS, COSMOS) to provide redshifts (spectroscopic and photometric)
for a large fraction of this sample (65\%). {\tt GALAPAGOS} provides
quantitative galaxy morphologies derived from {\tt GALFIT}
\citep{pen02}. We use single \citet{ser1968} models for all sources
detected by {\tt SExtractor} \citep{ber96}. In crowded fields we
simultaneously fit neighbors to avoid contamination by nearby
sources. For every source in the ACS-GC catalog we provide a color
image, a {\tt GALFIT} residual image, and an atlas image, which
combines all of the secondary data such as key morphological,
photometric, and redshift information into a single file. Refer to
Barden et al. (in prep.) or \citet{hau07} for a full description
of the {\tt GALAPAGOS} code and refer to Griffith et al. (in prep.)
for a full description of the ACS-GC.

% FIGURE 1
\begin{figure*}[!t]
\begin{center}
\begin{tabular}{c}
\includegraphics[scale=0.25,angle=0]{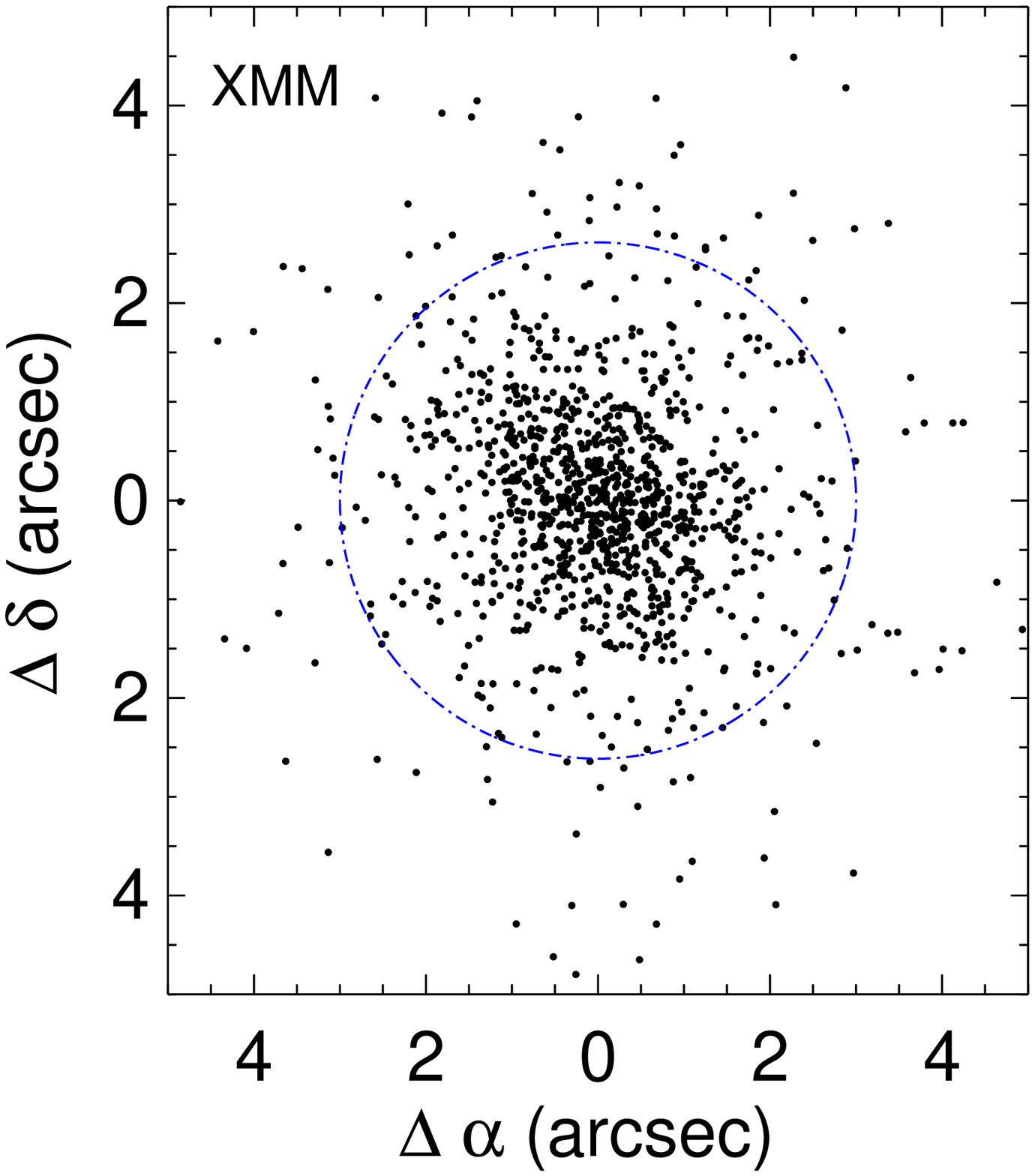}
\includegraphics[scale=0.25,angle=0]{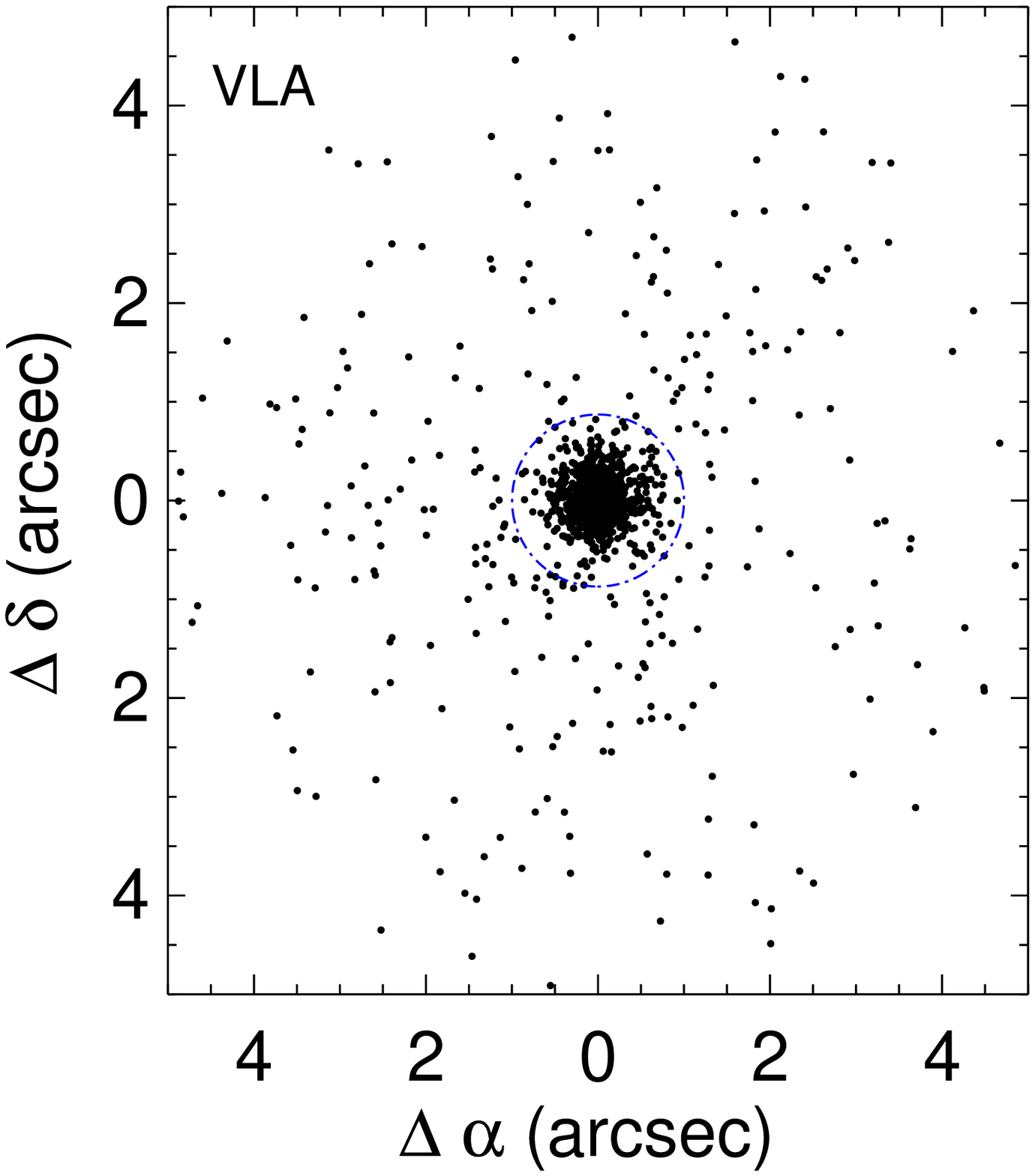}
\includegraphics[scale=0.25,angle=0]{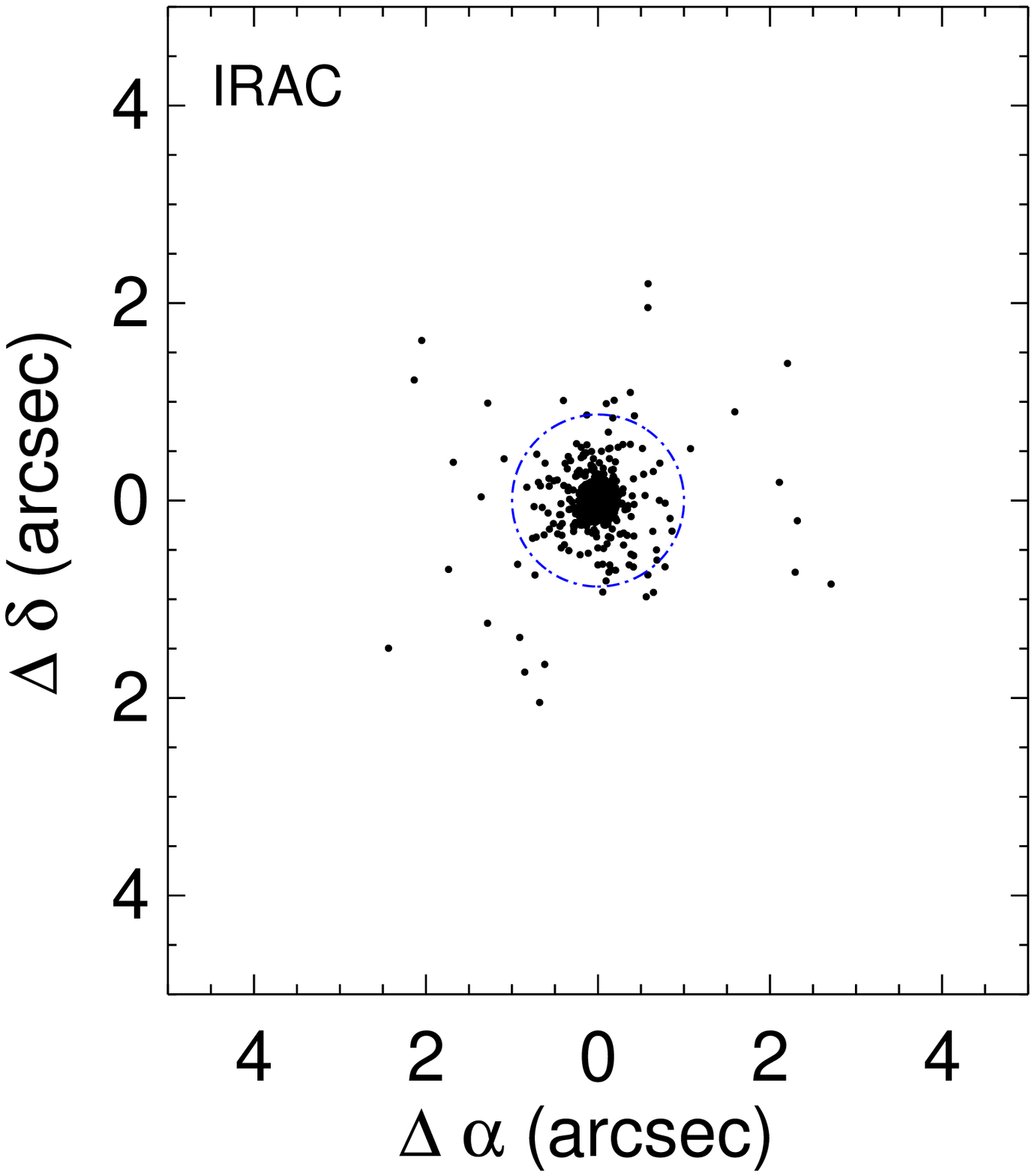}
\end{tabular}
\end{center}
\caption[CAPTION]{\label{fig:1} Positional offsets between nearest
ACS source to AGN candidates identified by {\it XMM-Newton} (left),
VLA (middle) and IRAC (right).  The final search radius used to
match catalogs is shown by the blue dashed circles:  for {\it{XMM}}-ACS
we use a 3\arcsec\ match radius, while both VLA-ACS and IRAC-ACS
use a 1\arcsec\ match radius.  Mean astrometric offsets between
catalogs derived from the test region have been applied (see
Table~1).}
\end{figure*}

\section{Cross-Matching Catalogs}

Optimal usage of these various data sets requires cross-matching
sources across the multi-wavelength data.  In order to determine
the best matching radii for the various COSMOS imaging data sets,
we created a sub-region, or test area, of the full survey which was
imaged by {\it Hubble}, {\it Spitzer}, {\it XMM-Newton}, and VLA.
We initially used very conservative (e.g., large) radii to match
the {\it Spitzer}, {\it XMM-Newton}, and VLA samples to the {\it
Hubble} imaging (5\arcsec, 5\arcsec\ and 3\arcsec, respectively).
We summarize the test area results in Table~1 ($N_{\rm test}$
column):  of 1161 {\it XMM-Newton} sources in the test area, we
matched 1117 (96\%) to ACS-GC sources within 5\arcsec; of 1534 VLA
sources ($S_{1.4} \ge 0.1$ mJy) in the test area, we matched 1440
(94\%) to ACS-GC sources within 5\arcsec; and we matched all 881
IRAC sources ([5.8] $<$ 16.5, Vega) in the test area to ACS-GC
sources within 3\arcsec.  We visually inspected those sources which
did not have an ACS-GC counterpart and found that most had faint
optical counterparts below the ACS-GC detection threshold.

Motivated by our cross-matching results (Figure~1) we use a 3\arcsec\
search  radius for the final {\it XMM}-ACS matching and a 1\arcsec\
search radius for the final VLA-ACS and IRAC-ACS matching.  These
final matchings also apply the mean astrometric offsets derived
from the test area, typically $\le 0.2\arcsec$, between the AGN
selection catalog and the ACS-GC (see Table~1).  The 1\arcsec\
search radius for the IRAC-ACS and VLA-ACS matching yields a 4.4\%
spurious detection rate at the full depth of the ACS-GC catalog.
The {\it XMM}-ACS matching, which required a larger matching radius,
yields a 24.5\% spurious identification for $I814 \le 24.5$, which
corresponds to the depth at which $\sim$90\% of the X-ray sources
have counterparts.  The spurious identification rate, of course,
is lower for brighter optical sources, with the {\it XMM}-ACS
spurious identification rate dropping to less then 5\% at $I814 \le
22.5$, the magnitude limit considered for the visual morphological
analysis discussed in \S 4.2.

% TABLE 1
\begin{deluxetable}{cccccc}
\tablewidth{0pt}
\tablecaption{Cross matching results.}
\tablehead{
\colhead{Sample} &
\colhead{$N_{\rm tot}$} &
\colhead{$N_{\rm test}$} &
\colhead{$N_{\rm final}$} &
\colhead{$\Delta \alpha$ (\arcsec)} &
\colhead{$\Delta \delta$ (\arcsec)}}
\startdata
Radio  & 2901 & 1440/1534 (94\%)& 372/602 (62\%) &$-0.065 \pm 0.159$ & $+0.094 \pm 0.167$ \\
X-ray  & 1887 & 1117/1161 (96\%) & 1360/1887 (72\%) & $-0.119 \pm1.011$ & $-0.118 \pm 0.884$ \\
Mid-IR & 1278 & 881/881 (100\%) & 1238/1278 (97\%) & $-0.105 \pm 0.103$ & $+0.044 \pm0.137$ \\
\enddata
\tablecomments{$N_{\rm tot}$ is the total number of sources in the
public catalogs, $N_{\rm test}$ lists the total and matched number
of sources in the test region, and $N_{\rm final}$ gives the total
and matched number of sources in our main AGN samples.  The final
two columns list the mean positional offsets derived from the test
region so that each catalog matches the ACS astrometric frame.}
\end{deluxetable}

\subsection{VLA-ACS}

We construct two samples of radio AGN candidates in the COSMOS
field. VLA1 consists of brighter sources, with flux densities greater
than or equal to 1 mJy ($S_{1.4} \ge 1.0$ mJy).  VLA2 consists of
fainter sources, with flux densities of $0.3 \le S_{1.4} < 1.0$
mJy. These cuts were motivated by \cite{sey08} who, based on a
detailed analysis of deep multiwavelength follow-up of faint radio
sources in the 13$^H$ {\it XMM-Newton}/{\it Chandra} Deep Field
Survey, show that virtually every 1.4~GHz radio source brighter
than 1~mJy and the majority ($\approx 70\%$) of sources brighter
than 0.3~mJy are AGN and not star-forming galaxies. The VLA catalog
contains 602 sources with $S_{1.4} \ge 0.3$ mJy. We cross match
these sources to the entire ACS-GC catalog (320,293) and find 372
(62\%) have optical counterparts within 1\arcsec.  This relatively
low counterpart fraction is due to a combination of the radio
field-of-view being larger than the ACS field-of-view, as well as
radio sources with complex radio morphologies.  For example,
approximately 30\% of FIRST sources have resolved structure on
scales of $2 - 30\arcsec$ \citep{bec95}, with the classic example
being a double-lobed radio source comprised of two radio-emitting
lobes separated by up to a few Mpc straddling a host galaxy.  Our
simple positional matching scheme will find no (correct) identification
for such a configuration and thus is likely responsible for our low
identification fraction for the radio-selected AGN.  There are 124
and 248 galaxies in the VLA1 and VLA2 samples, respectively.

\subsection{{\it XMM}-ACS}

The {\it XMM-Newton} catalog contains 1887 sources, of which 1387
(72\%) have ACS counterparts within 3\arcsec. We identify 27 sources
which are clearly Galactic stars ($I814 \le 15.5$ and CLASS\_STAR
$\ge$ 0.9; see \S3.4). Although this does not identify all of the
Galactic stars in the {\it XMM-Newton} catalog, it does flag the
brightest, least ambiguous Galactic sources. We thus have a total
of 1360 {\it XMM-Newton} sources with ACS counterparts that are not
bright, saturated stars.  We construct two samples of X-ray sources.
XMM1 consists of sources with soft (0.5-2.0 kev) X-ray fluxes
$S_{0.5-2.0} \ge 5.0\times 10^{-15}$ erg cm$^{-2}$ s$^{-1}$ and the
fainter sample, XMM2, consists of sources with  $S_{0.5-2.0} <
5.0\times 10^{-15}$ erg cm$^{-2}$ s$^{-1}$.  At the relatively
shallow depth of the {\it XMM-Newton} COSMOS survey, it is expected
that the vast majority of detected X-ray sources will be AGN
\citep[e.g.,][]{eck06}.  There are 338 and 1022 sources in XMM1 and
XMM2, respectively.

\subsection{IRAC-ACS}

We use the IRAC ``wedge" \citep{ste05} to select AGN candidates in
the S-COSMOS data. This selection relies on the fact that while
stellar populations fade longward of their 1.6 $\mu$m peak, luminous
AGNs exhibit red power-law emission throughout the mid-IR.  This
power-law, long-wavelength emission is relatively immune to gas and
dust obscuration, making mid-IR selection sensitive to both
obscured and unobscured luminous AGN \citep[e.g.,][]{lac04, ste05,
alo06, donley07}.  However, this AGN selection technique fails for
low-luminosity AGNs where the host galaxy contributes much of the
mid-IR flux \citep[e.g.,][]{donley07, don08, eck10}.  Note also
that for extremely deep mid-IR data, the AGN color selection methods
are contaminated by normal galaxies \citep[e.g.,][]{don08}, though
this should be a minimal effect for the bright mid-IR AGN samples
considered here.  In particular, \citet{eck10} perform an X-ray
stacking analysis of X-ray undetected, mid-IR-selected AGN from
relatively shallow IRAC data.  They find that such sources have,
on average, significantly brighter and harder X-ray spectra than
average mid-IR sources, consistent with the hypothesis that shallow
mid-IR color selection identifies a population of predominantly
obscured, high-luminosity AGN.

We find a total of 1278 AGN candidates which have [5.8] $<$ 16.5
(Vega), of which 1238 (97\%) sources have ACS counterparts within
1\arcsec. We construct two samples of IRAC sources.  IRAC1 consists
of the brighter mid-IR AGN candidates, selected to the $5\sigma$
depth of the original IRAC Shallow Survey \citep{eis04}, the survey
used to derive the empirical mid-IR criteria applied here \citep{ste05}.
IRAC2 consists of a fainter sample, selecting candidates not already
in IRAC1 but detected in all four IRAC bands down to the full $5
\sigma$ depth of the {\it Spitzer} Deep, Wide-Field Survey (SDWFS)
\citep{ash09}.  These criteria require robust ($5 \sigma$) detections
in all four IRAC bands.  However, for the typical power-law source
selected by the AGN wedge criteria, the detection limits are generally
driven by the $5.8~ \mu$m band and correspond to flux density limits
of $5.68\, \mu$Jy ([5.8] = 15.78, Vega) for IRAC1 and  $5.68 >
S_{5.8} > 2.93\, \mu$Jy (15.78 $<$ [5.8] $<$ 16.50, Vega) for IRAC2.
There are 590 and 648 sources in the IRAC1 and IRAC2 samples,
respectively.

\subsection{Overlap between the AGN samples}

These samples are not completely independent; sources classified
as AGN candidates in one band are often classified as AGN candidates
in another band.  Studying a sample of AGN in the Bo\"otes field,
\cite{hic09} find considerable overlap between their mid-IR and
X-ray samples, while the radio-selected AGNs are distinct.  We find
similar results here, summarized in Table~2. When considering all
AGNs, regardless of optical morphology, we find a modest overlap
between sources identified as IRAC and X-ray AGN:  we find $37\%$
of the IRAC AGN candidates have an {\it XMM-Newton} counterpart
while $34\%$ of the {\it XMM-Newton} sources are also identified
as AGN candidates from their mid-IR colors.  These trends were also
observed by \citet{hic09}, with 33\%/50\% cross identification for
their IRAC-X-ray and X-ray-IRAC samples, respectively.  We find
less overlap between the radio and X-ray/IRAC samples, 4-16\% for
both.  Only 23 sources are simultaneously identified as AGN using
all three selection methods.

The simplest morphological classification is star--galaxy separation,
or, rather, distinguishing unresolved sources (which might be
Galactic stars or quasars) from resolved sources (typically galaxies).
We use the {\tt SExtractor} CLASS\_STAR parameter to make this
separation, with unresolved sources defined to have CLASS\_STAR
$\ge 0.9$, and the remaining sources considered resolved.   We now
consider how this simple morphological criterion plays in with the
overlap between the AGN samples.

The 23 sources simultaneously identified using all three AGN selection
methods (e.g., X-ray bright, radio-loud AGN with mid-IR colors
indicative of an AGN) are approximately evenly divided between
optically extended sources (12) and optically unresolved sources
(11).  If we consider only those AGN with optically {\em extended}
morphologies, we find the overlap between the AGN selection techniques
is rather low:  only 23\% of optically resolved IRAC AGN candidates
have {\it XMM-Newton} counterparts, and only 23\% of optically
resolved {\it XMM-Newton} sources have mid-IR colors indicative of
AGN activity.  Considering the VLA sample, only 14\% of optically
resolved radio sources have {\it XMM-Newton} counterparts, and only
11\% of optically resolved radio sources have mid-IR colors indicative
of AGN activity.  This implies that to a large degree the various
AGN selection techiques identify unique classes of AGNs.

The optically {\em unresolved} AGN show a very different trend
altogether.  Considering just the AGN with point source optical
identifications, we find 71\% of the IRAC sources have {\it XMM-Newton}
counterparts and 57\% of the {\it XMM-Newton} sources have IRAC
counterparts.  We also find that virtually all (19/22, or 86\%) of
the optically unresolved radio sources have an IRAC counterpart and
50\% have an {\it XMM-Newton} counterpart.  This implies that these
sources represent similar classes of AGN, generally comprised of
unobscured quasars.  Note that we treat the AGN samples as independent
populations in much of the following analysis, identifying morphological
trends as a function of selection criterion and flux density.
However, the reader is cautioned to keep in mind the overlap between
the samples.

% TABLE 2
\begin{deluxetable}{cccccc}
\tablewidth{0pt}
\tablecaption{Overlap between AGN samples.}
\tablehead{
%\cutinhead{also identified in}
\colhead{Sample} &
\colhead{Total} &
\colhead{Radio} &
\colhead{X-ray}  &
\colhead{Mid-IR} &
\colhead{Radio$+$X-ray$+$Mid-IR}}
\startdata
Radio &  372 & \nodata  & 60 (16\%)  &  58 (16\%) & 23 (6\%) \\
X-ray & 1360 & 60 (4\%) & \nodata    & 463 (34\%) & 23 (2\%) \\
Mid-IR  & 1238 & 58 (5\%) & 463 (37\%) & \nodata    & 23 (2\%) \\
\cutinhead{optically extended AGNs (CLASS\_STAR $< 0.9$)}
Radio &  350 & \nodata  &  49 (14\%) &  39 (11\%) & 12 (3\%) \\
X-ray &  888 & 49 (5\%) & \nodata    & 194 (22\%) & 12 (1\%) \\
Mid-IR  &  857 & 39 (5\%) & 194 (23\%) & \nodata    & 12 (1\%) \\
\cutinhead{optically unresolved AGNs (CLASS\_STAR $\ge 0.9$)}
Radio &   22 & \nodata  &  11 (50\%) &  19 (86\%) & 11 (50\%) \\
X-ray &  472 & 11 (2\%) & \nodata    & 269 (57\%) & 11 (2\%) \\
Mid-IR  &  381 & 19 (5\%) & 269 (71\%) & \nodata    & 11 (3\%) \\
\enddata
\tablecomments{The final column, Radio$+$X-ray$+$Mid-IR, refers to
AGN candidates identified by all three selection techniques.}
\end{deluxetable}

\section{Morphologies of Radio-, X-ray-, and Mid-IR-Selected AGN}

In this section we describe two methods for investigating how host
galaxy morphology correlates with AGN type. In $\S$4.1 we use the
S\'ersic index to classify galaxies and find a large fraction of
AGN hosts have high S\'ersic values ($n \ge 2.5$), implying populations
heavily dominated by bulge-dominated or early-type morphologies.
However, in $\S$4.2 we visually classify the brightest ($I814 \le
22.5$) galaxies in the sample and find that the X-ray and mid-IR
populations are not dominated by early-type galaxies, but by disk
galaxies. These results motivate the need for more complex quantitative
methods for understanding these populations of galaxies.  At the
present time when the sample sizes are of order a few thousand,
visual morphologies are still manageable.  But as all-sky,
high-resolution imaging surveys come online in the future, we will
need to develop reliable and consistent algorithms for classifying
these galaxies.

We note that surface brightness dimming could be an issue for the
morphological assessments, making faint disks difficult to identify.
However, this is not expected to be a significant concern, especially
for the brighter sample with visual morphological classifications.
\citet{gab09} have done extensive simulations of how well morphological
parameters are recovered in the presence of noise.  These simulations,
done as a function of galaxy mean surface brightness, show that
morphological parameters for galaxies with $\mu \leq 23$~mag~arcsec$^{-2}$
are recovered quite reliably and with no systematic offsets; e.g.,
S{\'e}rsic indices are recovered with $\sigma_n$(disk galaxies)
$\simlt 0.1$ and $\sigma_n$(spheroids) $\approx 1$, and thus are
unlikely to change the morphological classification of sources.  At
$\mu = 24$~mag~arcsec$^{-2}$, these uncertainties are slightly
higher, $\sigma_n$(disk galaxies) $\approx 0.3$ and $\sigma_n$(spheroids)
$\approx 2$.  For the $I814 < 22.5$ sample, 79\% (88\%) of AGN
candidates have $\mu \leq 23$~mag~arcsec$^{-2}$ ($\mu \leq
24$~mag~arcsec$^{-2}$).  These brighter sources will largely allow
for robust morphological classifications.  Considering the full
sample of AGN candidates, 63\% (76\%) of sources have $\mu \leq
23$~mag~arcsec$^{-2}$ ($\mu \leq 24$~mag~arcsec$^{-2}$), with between
48\% (IRAC2) and 90\% (XMM1) of the various AGN subsamples having
mean surface brightnesses $\leq 23$~mag~arcsec$^{-2}$ and between
67\% (IRAC2) and 94\% (XMM1) having $\mu \leq 24$~mag~arcsec$^{-2}$.
Therefore, though we expect the faintest galaxies to have uncertain
morphological measurements, the brighter sources which dominate our
various samples should allow for accurate morphological assessments
and thus the statistical trends discussed below are expected to be
robust.

\subsection{Quantitative morphologies}

The final COSMOS ACS-GC catalog contains 320,293 sources, which we
use to quantitatively investigate the morphological characteristics
of the AGN hosts. The {\tt GALAPAGOS} analysis was performed prior
to the AGN morphology investigation discussed herein, and hence we
did not undertake any special procedures to account for potential
contamination from a nuclear point source.  In contrast, \citet{gab09}
analyzed 389 AGN candidates, simultaneously fitting an unresolved
component with the resolved galaxy component.  Using simulated
images, \citet{gab09} show that a nuclear point source can significantly
bias the measurements of structural parameters. In practice, however,
they find that only a small fraction of sources have a point source
contribution significant enough to adversely effect the quantitative
morphological parameters --- specifically, only 7\% of sources have
a point source $\le$ 3 mag fainter than the host galaxy.
\citet{Pierce:10} found that AGNs which contribute $\ge$ 20\% of
the galaxy light can bias morphological measurements, making such
systems appear bluer and more bulge dominated. Pierce et al. conclude
that bulge-dominated, E/S0/Sa, and early-type morphology classifications
are accurate for AGN host galaxies, unless the AGN manifests itself
as a well-defined point source which can be flagged by inspecting
the color images. With this in mind, this section discusses the
automated, quantative morphological results for the AGN samples,
while in $\S$4.2 we also perform visual morphological classifications
for the brightest galaxies in the sample, $I814 \le 22.5$.

% TABLE 3
\begin{deluxetable}{ccccccc}
\tablewidth{0pt}
\tablecaption{Quantitative AGN morphologies.}
\tablehead{
\colhead{Sample} &
\colhead{IRAC1} &
\colhead{IRAC2} &
\colhead{XMM1} &
\colhead{XMM2} &
\colhead{VLA1} &
\colhead{VLA2}}
\startdata
$N_{\rm tot}$  &   590 &           648 &           338 &        1022 &      124 &     248 \\
point source   &  257 (44\%) &       124 (19\%)      &  189 (56\%) & 283 (28\%) &   8 (6\%)  &  14 (6\%) \\
extended       &  333 (56\%) &       524 (81\%)      &  149 (44\%) & 739 (72\%) & 116 (94\% )& 234 (94\%) \\
\cutinhead{optically extended AGNs (CLASS\_STAR $< 0.9$)}
$n=0.2$             &   4  (1\%) &  13  (3\%) &  1  (1\%) &  12  (2\%) &  1  (1\%) &   5  (2\%) \\
$0.2 < n <$ 1.5     &  56 (17\%) & 146 (28\%) & 17 (11\%) & 120 (16\%) &  8  (7\%) &  49 (21\%) \\
$1.5 \le n \le$ 2.5 &  35 (11\%) &  66 (13\%) &  6  (4\%) &  78 (11\%) &  5  (4\%) &  28 (12\%) \\
$2.5 < n <$ 8       & 135 (41\%) & 212 (41\%) & 52 (35\%) & 326 (44\%) & 71 (61\%) & 112 (48\%) \\
$n=8$               & 103 (31\%) &  87 (17\%) & 73 (49\%) & 203 (28\%) & 32 (28\%) &  40 (17\%) \\
\enddata
\end{deluxetable}

We derive the following quantitative parameters for the AGN host
galaxies: half-light or effective radius ($r_e$), S\'ersic index
($n$), apparent magnitude (MAG\_BEST), and source stellarity index
(CLASS\_STAR).  The first two are measured with {\tt GALFIT} and
the latter two are measured with {\tt SExtractor}. The effective
radius is the radius to which one half of the total light is emitted
and is traditionally used to measure the size of a source. The
S\'ersic index is derived from the \citet{ser1968} profile and
describes how the intensity of a source varies with distance from
the center. The S\'ersic profile represents a general form that
encompasses both the exponential profile ($n=1$), characteristic
of disk galaxies, and the de Vaucouleurs profile ($n=4$), characteristic
of elliptical galaxies.  We restrict the S\'ersic index to be in
the range $0.2 \le n \le 8$.  The stellarity index, which ranges
from zero to one, separates compact sources (stars and quasars;
CLASS\_STAR $\ge$ 0.9) from extended galaxies (CLASS\_STAR $<$ 0.9).
Note that for stellar sources and compact galaxies, morphological
parameters such as $r_e$ and $n$ are generally not reliable due to
the sensitivity of the results to the assumed PSF.

% TABLE 4
\begin{deluxetable}{ccccccc}
\tablewidth{0pt}
\tablecaption{Quantitative AGN morphologies ($I814 \le 22.5$).}
\tablehead{
\colhead{Sample} &
\colhead{IRAC1} &
\colhead{IRAC2} &
\colhead{XMM1} &
\colhead{XMM2} &
\colhead{VLA1} &
\colhead{VLA2}}
\startdata
$N_{\rm tot}$  &  444 &        238 &         280 &        458 &      81 &     188 \\
point source   &  228 (51\%) &  70 (29\%) &  172 (61\%) & 150 (33\%) &  7 (9\%) & 11 (6\%)\\
extended       &  216 (49\%) & 168 (71\%) &  108 (39\%) & 308 (67\%) & 74 (91\%) & 177 (94\%)\\
\cutinhead{optically extended AGNs (CLASS\_STAR $<$ 0.9)}
$n=0.2$             &  1  (1\%) &  4  (2\%) &  1  (1\%) &   1  (0\%) &  1  (1\%) &  1  (1\%)\\
$0.2 < n <$ 1.5     & 40 (19\%) & 58 (35\%) &  9  (8\%) &  39 (13\%) &  4  (5\%) & 44 (25\%)\\
$1.5 \le n \le$ 2.5 & 25 (12\%) & 17 (10\%) &  5  (5\%) &  29  (9\%) &  2  (3\%) & 26 (15\%)\\
$2.5 < n <$ 8       & 81 (38\%) & 71 (42\%) & 36 (33\%) & 175 (57\%) & 54 (73\%) & 88 (50\%)\\
$n=8$               & 69 (32\%) & 18 (11\%) & 58 (54\%) &  65 (21\%) & 14 (19\%) & 19 (11\%)\\
\enddata
\end{deluxetable}

% FIGURE 2
\begin{figure*}[!t]
\begin{center}
\begin{tabular}{c}
\includegraphics[scale=0.75,angle=90]{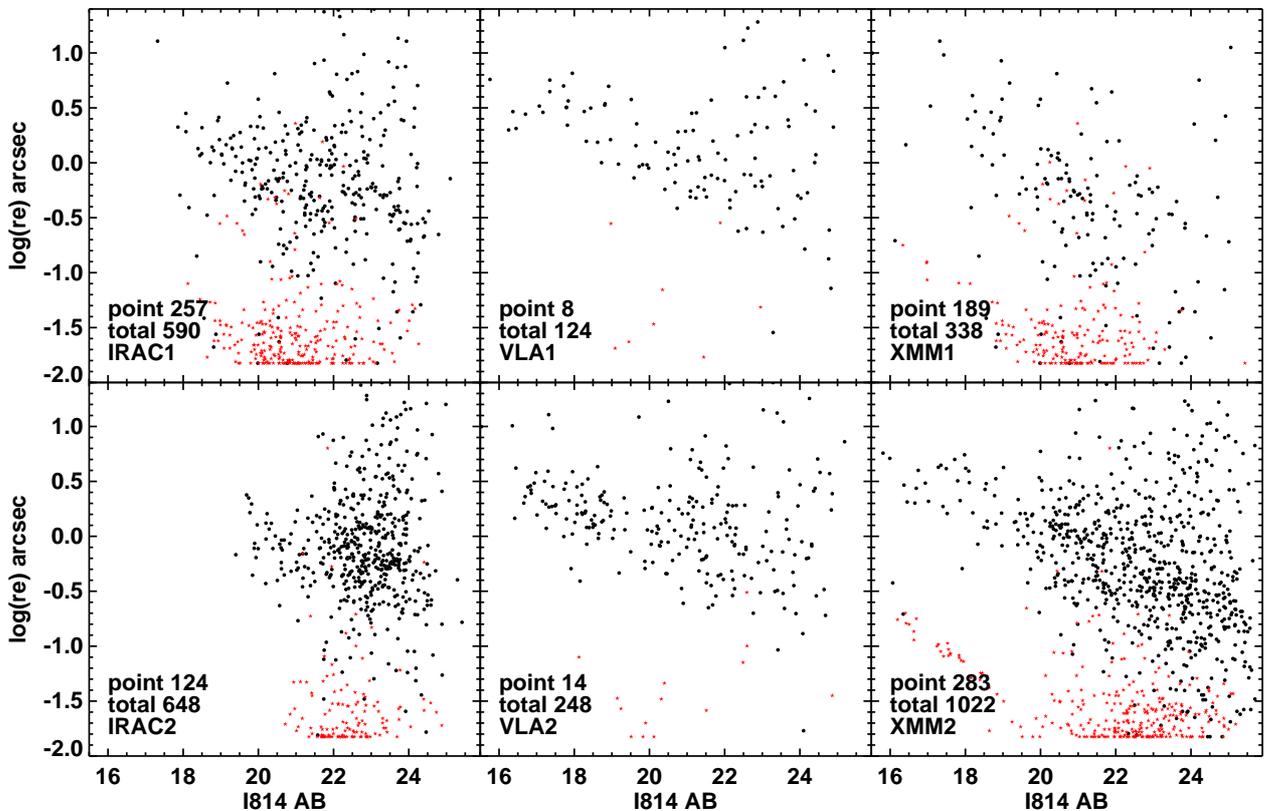}
\end{tabular}
\end{center}
\caption[CAPTION]{\label{fig:2} Half-light radius, $r_e$, versus
{\tt SExtractor} $I814$ MAG\_BEST. Red stars represent unresolved
sources (e.g., {\tt SExtractor} CLASS\_STAR $\ge$ 0.9). Black points
represent sources with {\tt SExtractor} CLASS\_STAR $<$ 0.9.  Text
in each panel give the total number of sources plotted as well as
the number which are optical unresolved (``point'').}
\end{figure*}

% FIGURE 3
\begin{figure*}[!t]
\begin{center}
\begin{tabular}{c}
\includegraphics[scale=0.75,angle=90]{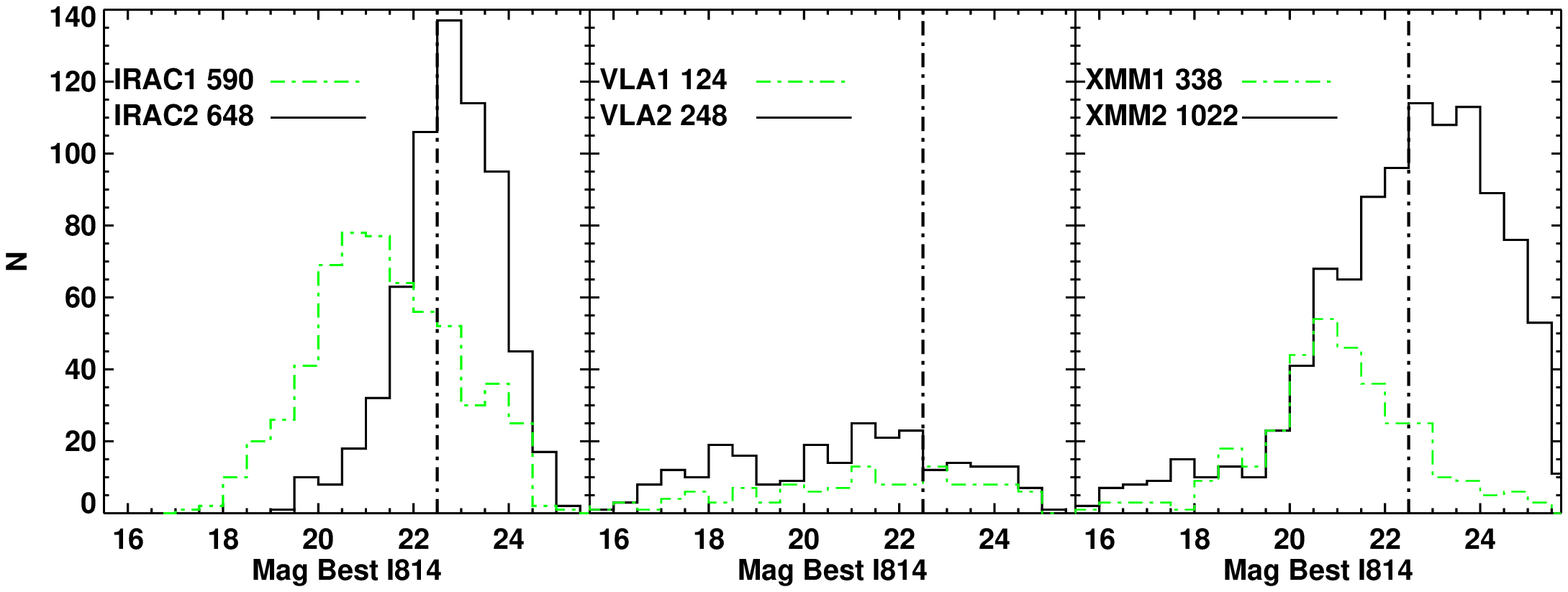}\\
\includegraphics[scale=0.75,angle=90]{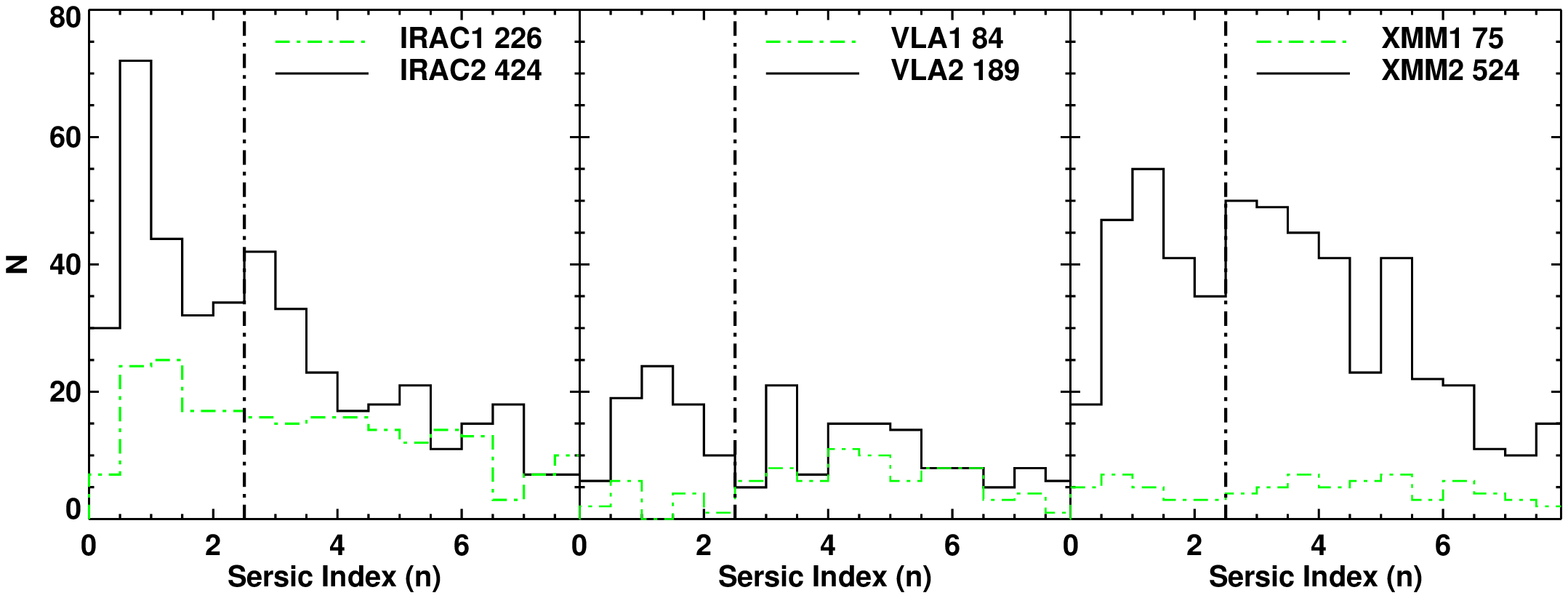}
\end{tabular}
\end{center}
\caption[CAPTION]{\label{fig:3} {\bf Top:} $I814$ MAG\_BEST
distribution for all sources in the samples, including both resolved
and unresolved hosts.  The vertical dot-dashed line illustrates the
magnitude limit to which visual morphologies were performed ($I814
= 22.5$; \S 4.2).  {\bf Bottom:} S\'ersic index ($n$) distribution
for spatially-resolved hosts with $0.2 < n < 8$. The vertical
dot-dashed line is the traditional dividing line between early-type,
bulge-dominated galaxies ($n > 2.5$) and late-type, disk galaxies
($n \le 2.5$).}
\end{figure*}

IRAC1 has a roughly equal number of point sources and extended
sources (44/56\%) while IRAC2, the fainter mid-IR AGN sample, is
dominated by extended sources (81\%). We find a similar trend for
the X-ray sources with a roughly equal numbers of point sources and
extended sources (56/44\%) in XMM1 while XMM2 is dominated by
extended sources (72\%). The radio sample is quite different, with
extended sources dominating (94\%) both subsamples.  Figure~2 plots
size versus $I814$ magnitude for the six samples.  We observe that
the VLA samples have overall larger sizes as well as brighter
magnitudes as compared to the IRAC and X-ray samples. XMM2 also
appears to have more faint optical counterparts, potentially due
to contamination from spurious sources, which are expected to
contribute $\sim$ 25\% of the identifications at $I814 \sim$ 24.5.

The size-magnitude relationship has traditionally been used to
distinguish extended galaxies (e.g., normal and low surface brightness
galaxies), compact galaxies and stellar-like objects.  Quasars,
stars, and compact galaxies populate the stellar locus and have
high stellarity values.  Figure 2 shows several surprising examples
of sources with high stellarity that are also spatially extended.
In order to understand these sources, we visually inspected a sample
of 39 sources with CLASS\_STAR $\ge$ 0.9 and $r_e \ge 0.1\arcsec$.
A majority of these sources have large residuals when the best fit
morphological model is subtracted, implying  that the galaxy model
and the true galaxy image are not consistent and that the galaxy
structural parameters are thus unreliable.  A few (9 or 23\%) of
these sources are extended galaxies hosting a bright point source
(e.g., \S 5.1).  For the following analysis, we remove all sources
with CLASS\_STAR $\ge$ 0.9 due to the uncertainties in their
structural parameters.

We divide the extended sources into five bins using the S\'ersic
index, with two of these bins being the boundary of our S\'ersic
models ($n=0.2$ and $n=8$; see Tables~3 and 4, and Fig.~3).  The
other three bins are $0.2 < n < 1.5$, comprised of late-type and
spiral galaxies, $1.5 \le n \le 2.5$, comprised of galaxies which
have blended morphologies (i.e., bulge + disk), and $2.5 < n < 8$,
generally comprised of elliptical or early-type galaxies. We visually
inspected the 32 spatially-extended sources which have $n=0.2$ and
find that these sources are very low surface brightness galaxies
with disturbed morphologies. Galaxies that have $n$ values which
reach the boundaries are generally removed from further analysis.
For $n=0.2$, this will not significantly affect the results as such
sources only contribute between 1\% and 2\% of the total samples.
Eliminating $n=8$ sources will certainly affect the overall statistics
for the samples, with between 17\% (IRAC2) and 49\% (XMM1) of the
spatially extended samples having $n=8$.  For the extreme example
of the XMM1 sample, fully 78\% of the sources are either unresolved
or have S\'ersic index $n=8$.

On the other hand, were we to we combine the $n=8$ sources with the
$n > 2.5$ sources, we would find a misleading overabundance of
early-type morphologies.  AGN host galaxies comprise a very broad
range of morphological types ranging from morphologically ``normal"
galaxies which include elliptical and spiral galaxies to merging
systems with complicated structural components.  Some galaxy
structures are not easily fit with single component models and hence
return spurious results; these galaxies must be classified using
different methods.  We see that even when $n=8$ host galaxies are
considered distinct, the AGN hosts are still often classified as
bulge-dominated galaxies.  The radio samples have the largest such
percentages, with VLA1 (VLA2) comprised of 61\% (48\%) early-type
host galaxies ($2.5 < n < 8$).  The mid-IR and X-ray AGN samples
have $\sim$ 40\% bulge-dominated galaxies for both the faint and
bright samples.  Late-type morphologies ($0.2 < n < 1.5$) range
from 7\% for the VLA1 sample to 28\% for the IRAC2 sample.  Blended
galaxies ($1.5 \le n \le 2.5$) constitute $\sim$ 10\% of all of the
samples.

Figure 3 (top) shows the $I814$ magnitude distribution for our six
samples; the brighter samples are shown by green dashed lines.  We
find that the optical flux densities roughly correlate with both
the X-ray and mid-IR fluxes:  XMM1 and IRAC1 optical magnitudes are
systematically brighter, with the peak of the distribution around
$I814 \sim 21$, while the XMM2 and IRAC2 are systematically fainter
by about two magnitudes.  The radio sample does not show any
correlation between the optical and radio flux densities, implying
that the radio-selected AGN must sample a broad distribution of
Eddington ratios.

\subsection{Visual morphologies}

We also measured visual morphologies for the 1290 (unique) AGN
candidates with $I814\le 22.5$. Visual morphologies were performed
on pseudocolor images constructed by P.  Capak (private communication)
using the ACS $I814$ data as an illumination map and deep Subaru
$B_J$, $r^+$, and $i^+$ images as a color map. To achieve this,
each Subaru image was divided by the average of the three Subaru
images and then multiplied by the ACS $I814$ image. This preserves
the flux ratio between images while replacing the overall illumination
pattern with the $I814$ data. Each image was then divided by
$\lambda^{2}$ to enhance the color difference between star-forming
and passive galaxies. The processed  $B_J$, $r^+$, and $i^+$ images
were then assigned to the blue, green, and red channels, respectively.
The resulting images have the high spatial resolution of the ACS
imaging but color gradients at ground-based resolution. Visual
classifications were performed independently by the two authors,
and our independent analysis agreed for the majority (84\%) of the
sources. For the remaining 210 sources where the initial independent
analysis did not initially agree, we reviewed the sources together
and assigned a consensus morphology. Such sources tended to be red
galaxies with S0 morphologies.  We split the sample into five
morphological types (see Figure 4): point sources, bulge-dominated
galaxies, disk galaxies, S0 galaxies, and other:

% FIGURE 4
\begin{figure*}[!t]
\begin{center}
\begin{tabular}{c}
\includegraphics[scale=0.3]{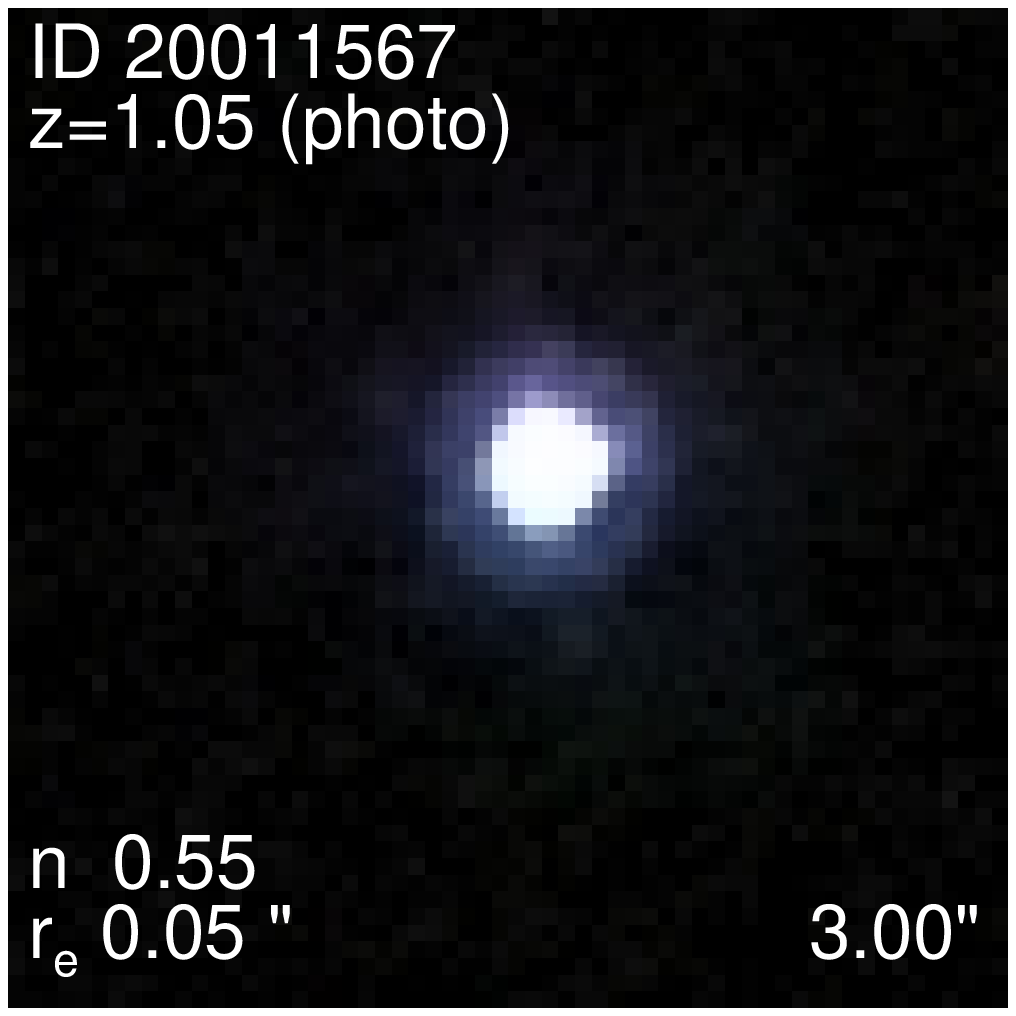}
\includegraphics[scale=0.3]{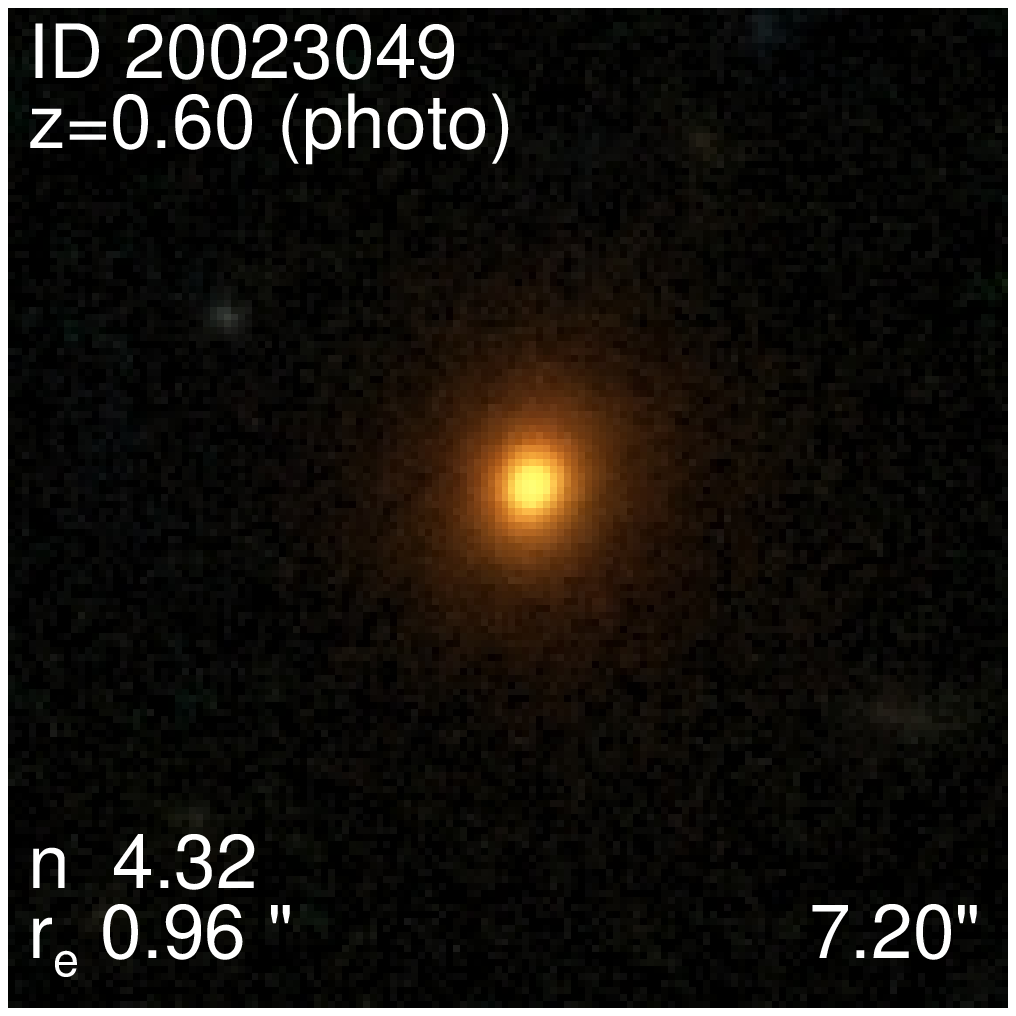}
\includegraphics[scale=0.3]{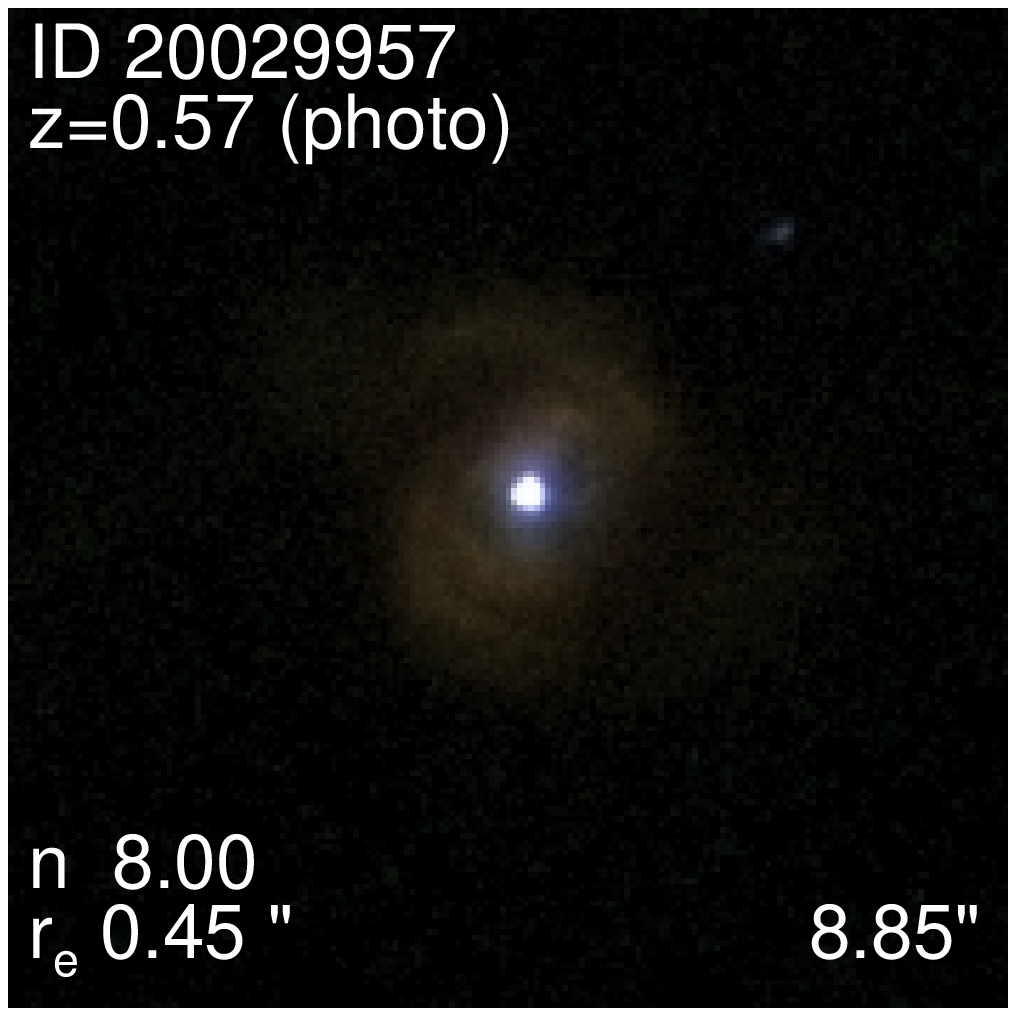}
\includegraphics[scale=0.3]{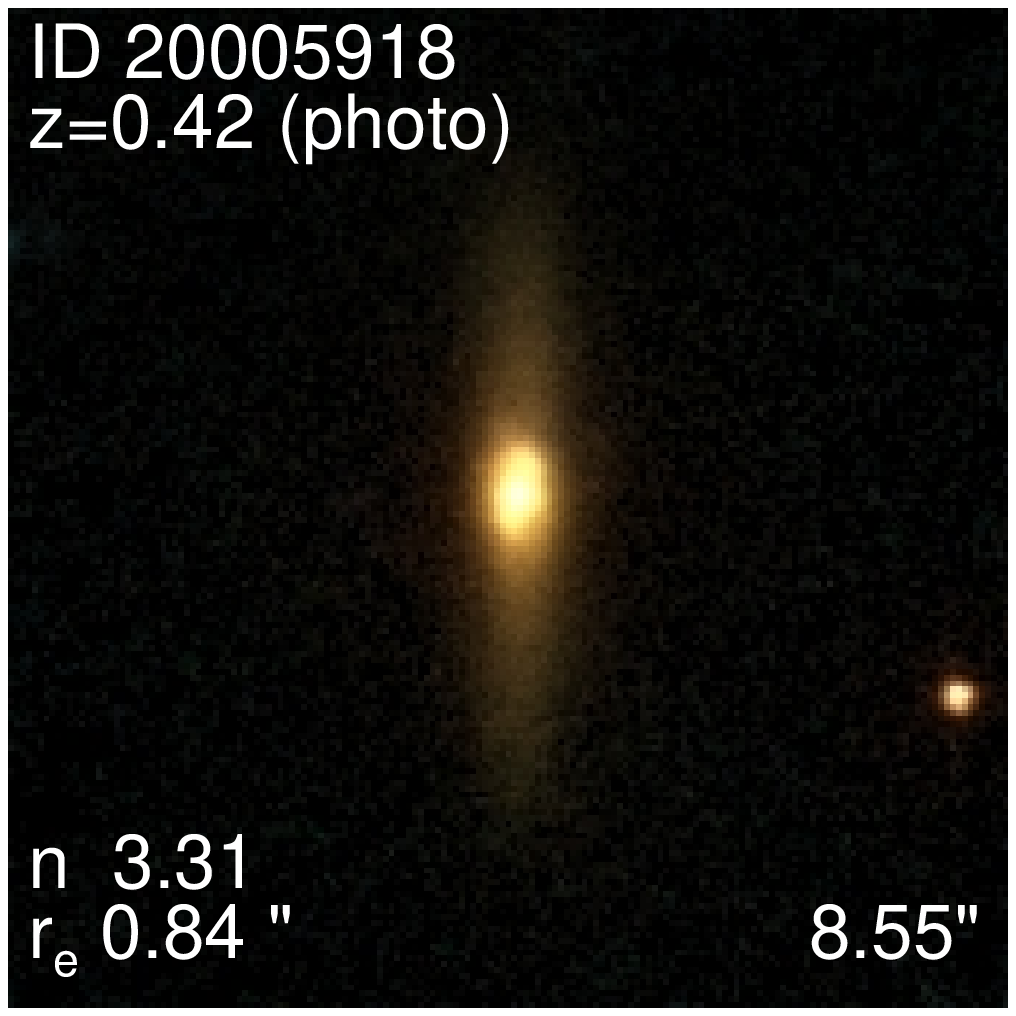}
\includegraphics[scale=0.3]{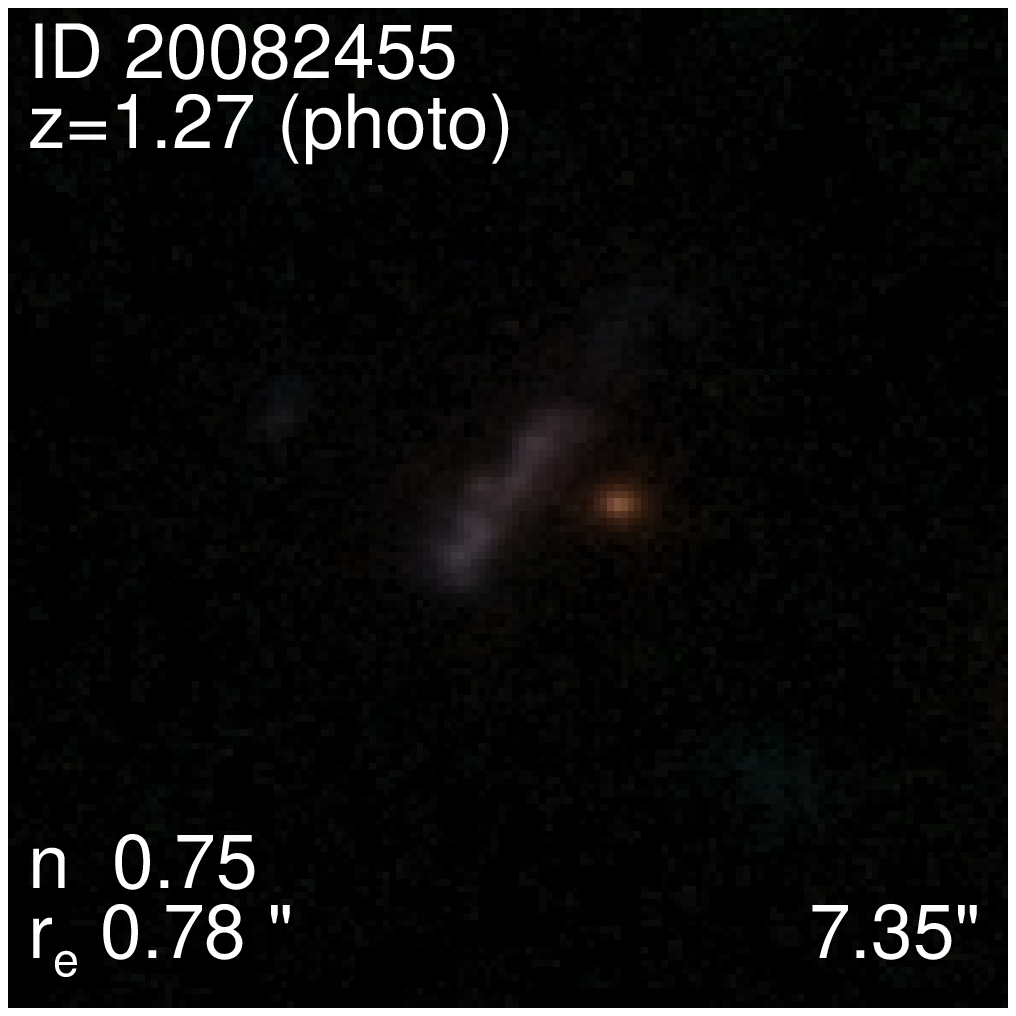}\\
\includegraphics[scale=0.3]{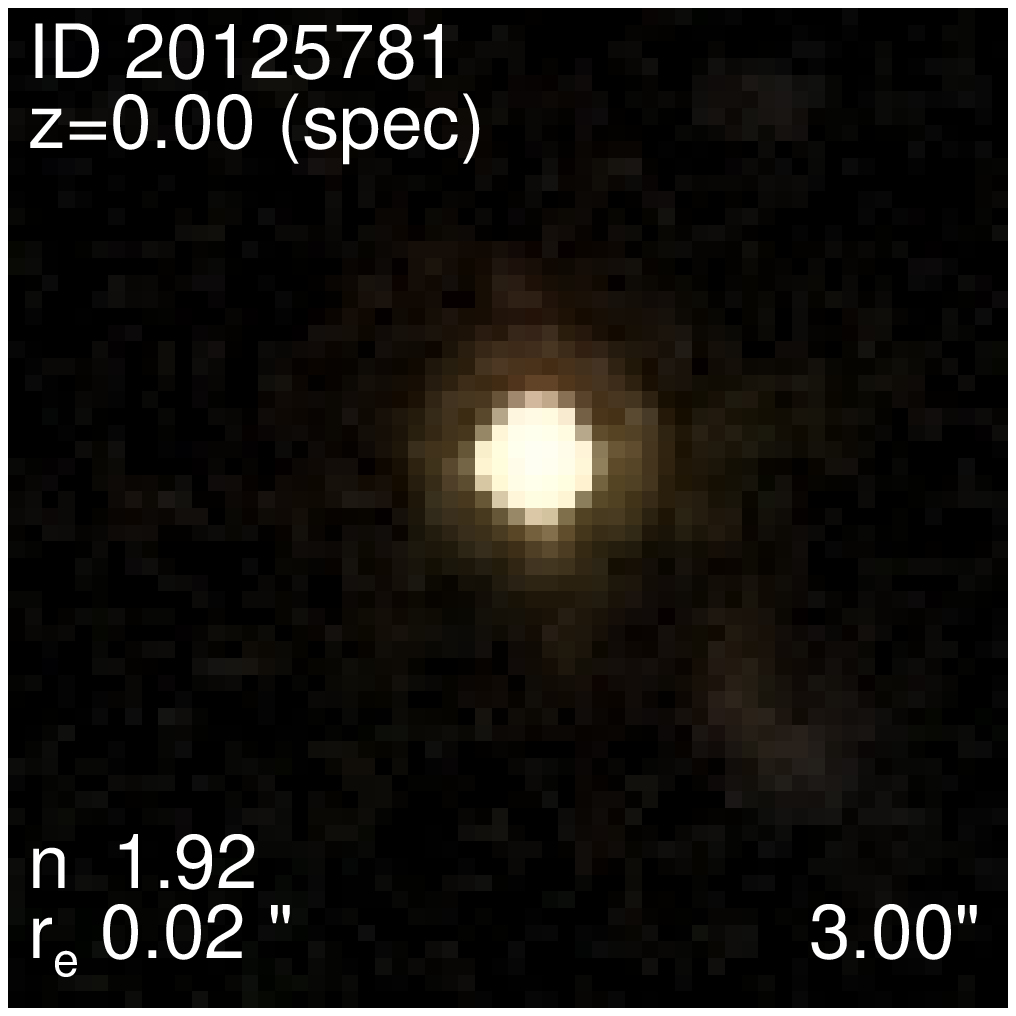}
\includegraphics[scale=0.3]{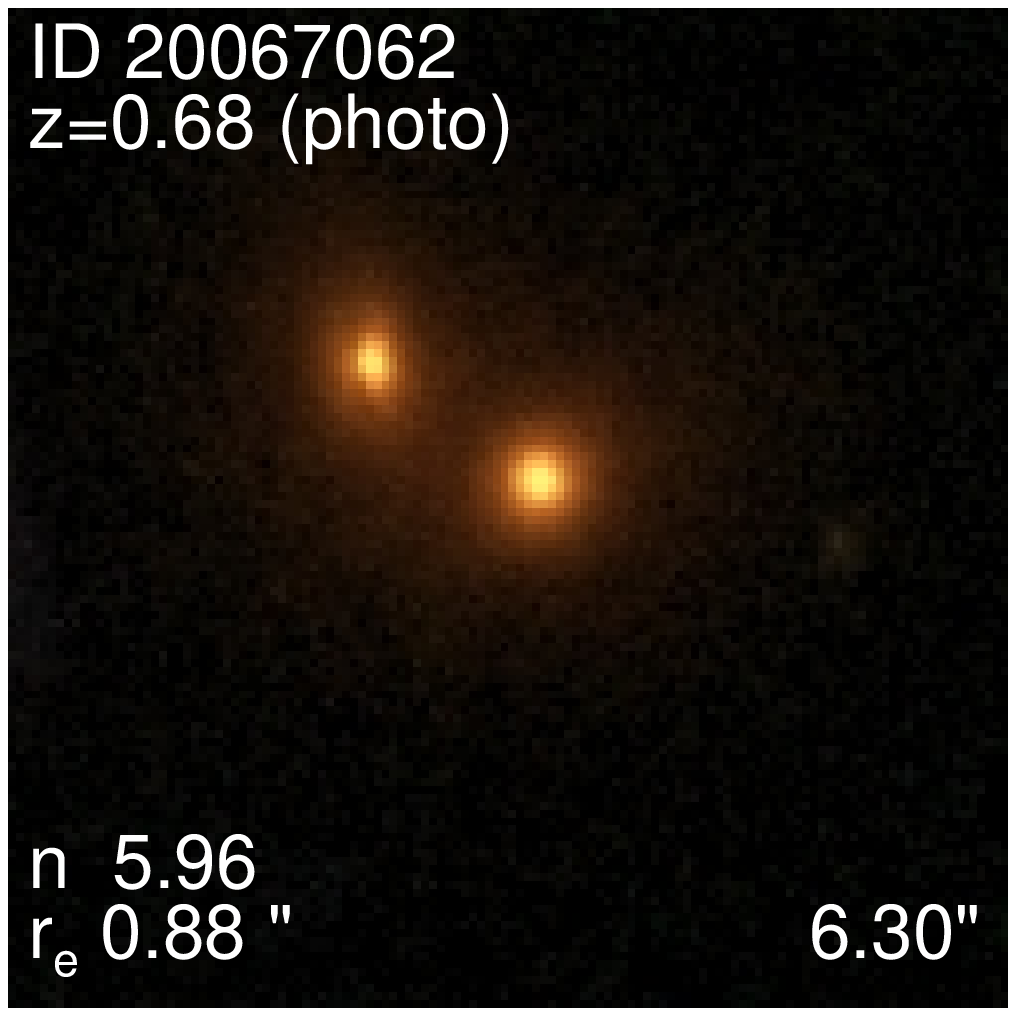}
\includegraphics[scale=0.3]{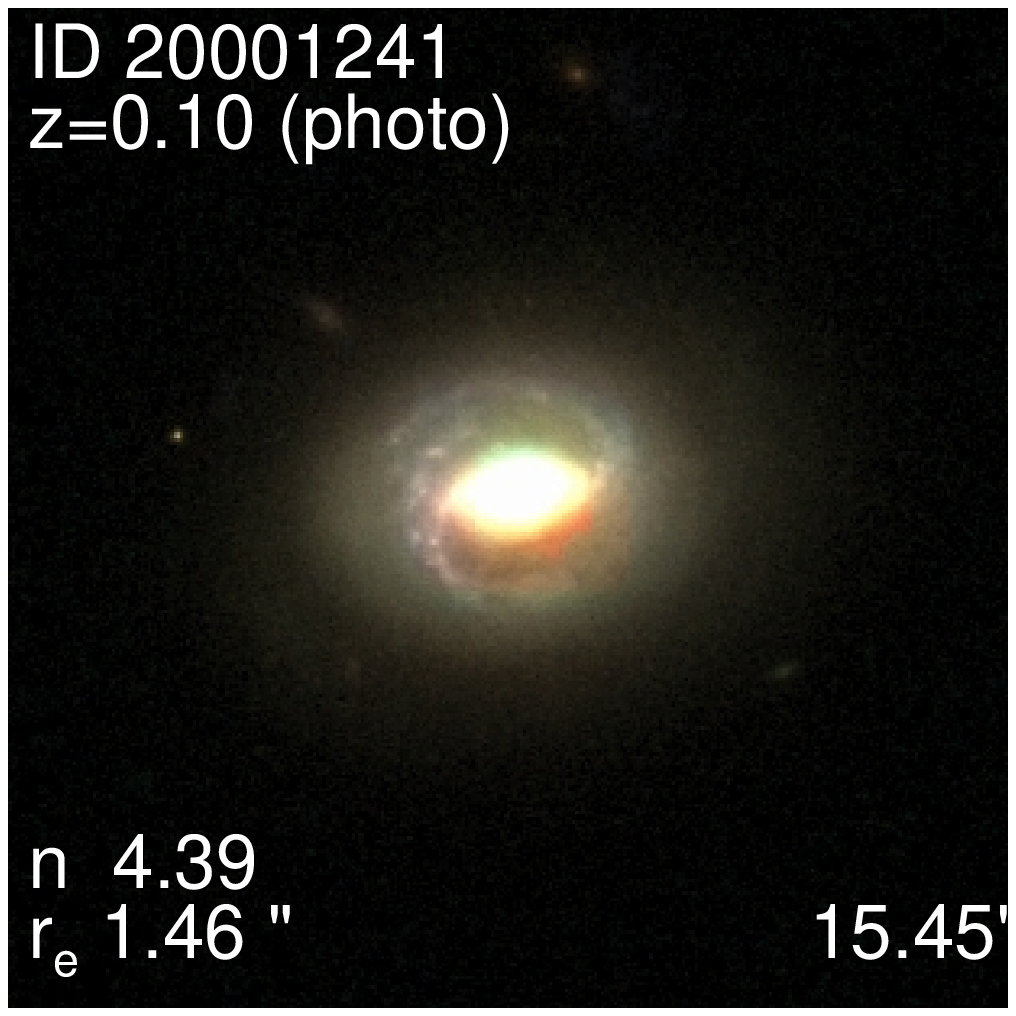}
\includegraphics[scale=0.3]{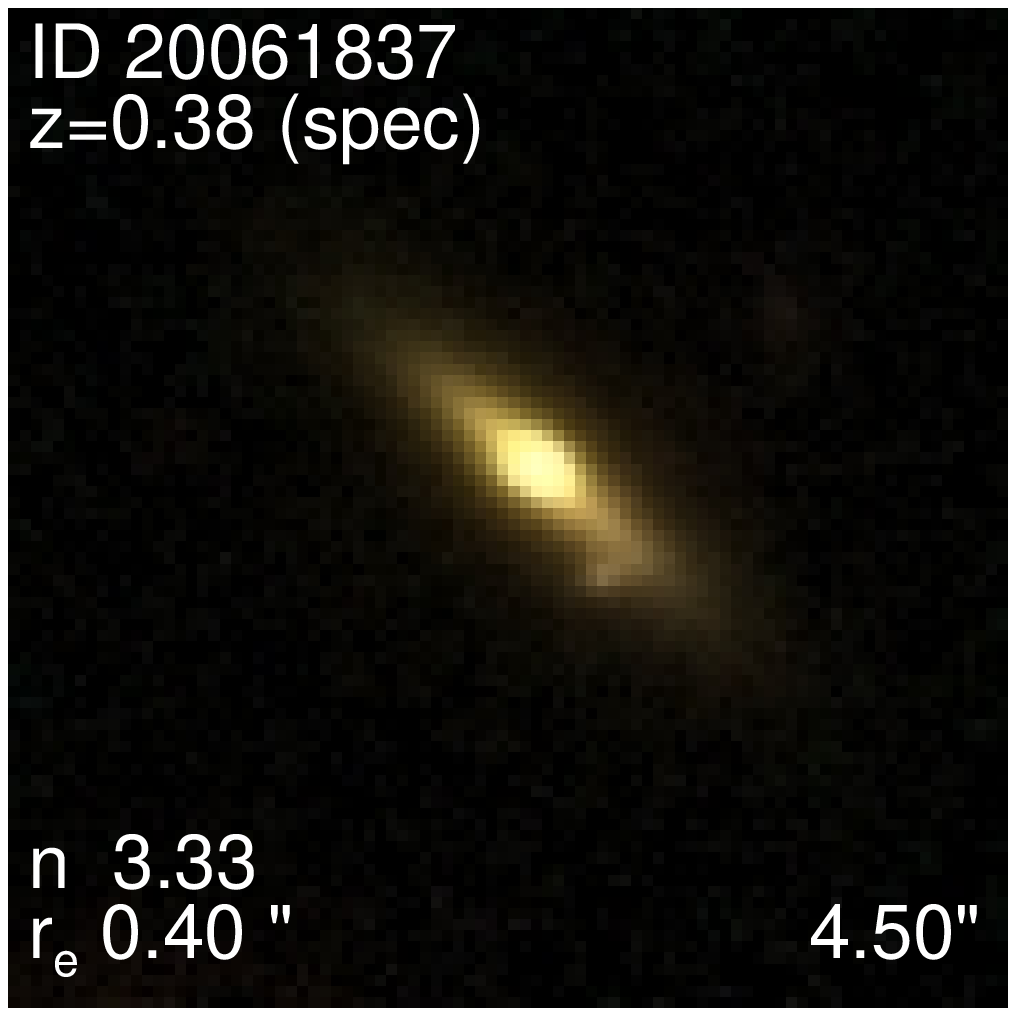}
\includegraphics[scale=0.3]{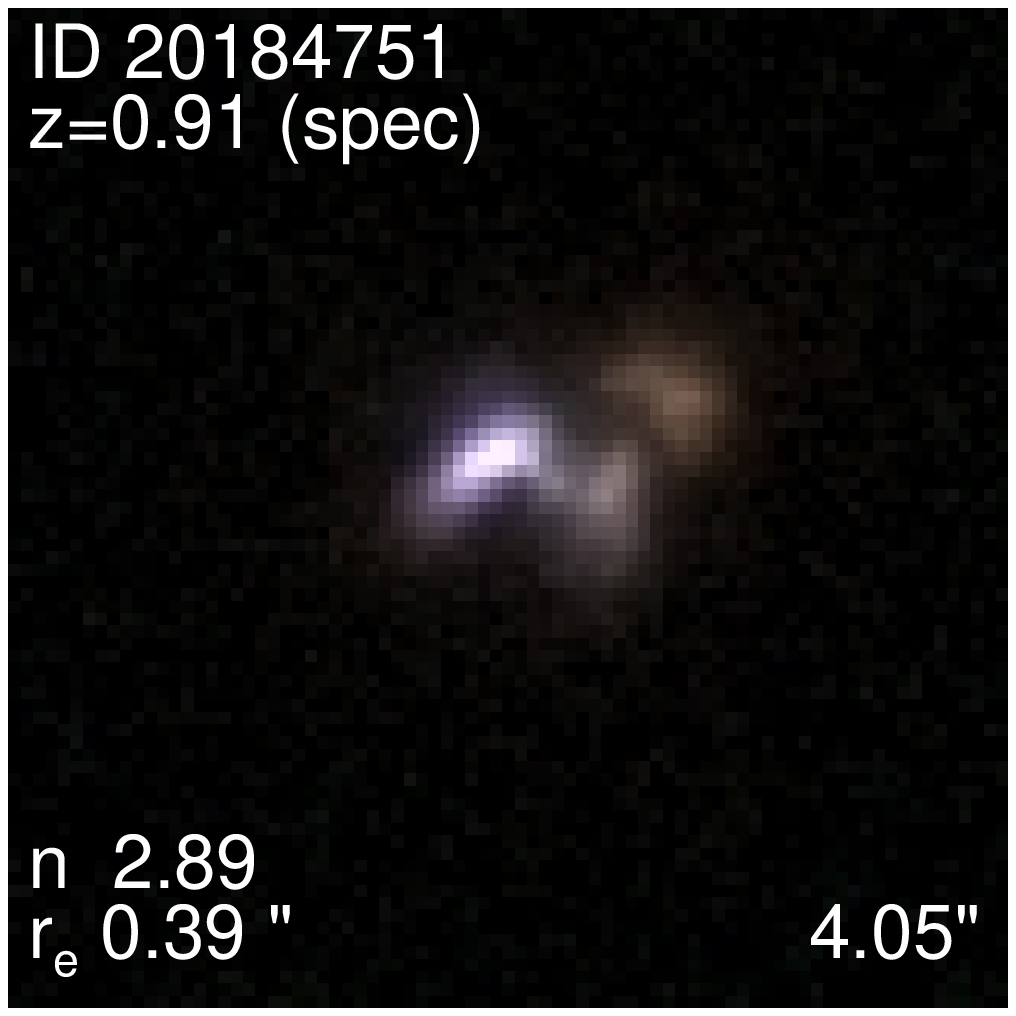}
\end{tabular}
\end{center}
\caption[CAPTION]{\label{fig:4} Representative visual morphological
classifications. Column 1: point sources; column 2: bulge-dominated
(e.g., early-type) galaxies; column 3: disk (e.g., late-type)
galaxies; column 4: S0 galaxies; and column 5: other.  We include
the ACS-GC object number (ID), the redshift (spectroscopic or
photometric), the S\'ersic index, the half-light radius, and the
postage stamp size (lower right).}
\end{figure*}

$\bullet$ {\it{point sources}} --- compact point sources with no
extended structural components. These include bright quasars and
Galactic stars;

$\bullet$ {\it{bulge-dominated galaxies}} --- elliptical galaxies
which are dominated by a bulge and lack a disk component;

$\bullet$ {\it{disk galaxies}} --- galaxies comprised of a wide
range of types (e.g., Sa thru Sc), all hosting a clear disk component
(though not necessarily disk-dominated);

$\bullet$ {\it{S0 galaxies}} --- early-type galaxies which have an
extended disk component and red colors (this classification has
some overlap with the disk classification and, in fact, is where
many of the independent morphological visual classifications
disagreed; however, S0 galaxies are distinguished by their lack of
spiral structure in the disk component, as well as their strong
bulge component); and

$\bullet$ {\it{other}} --- disturbed morphologies or morphologies
which cannot be easily classified.

% FIGURE 5
\begin{figure*}[!t]
\begin{center}
\begin{tabular}{c}
\includegraphics[scale=0.75,angle=90]{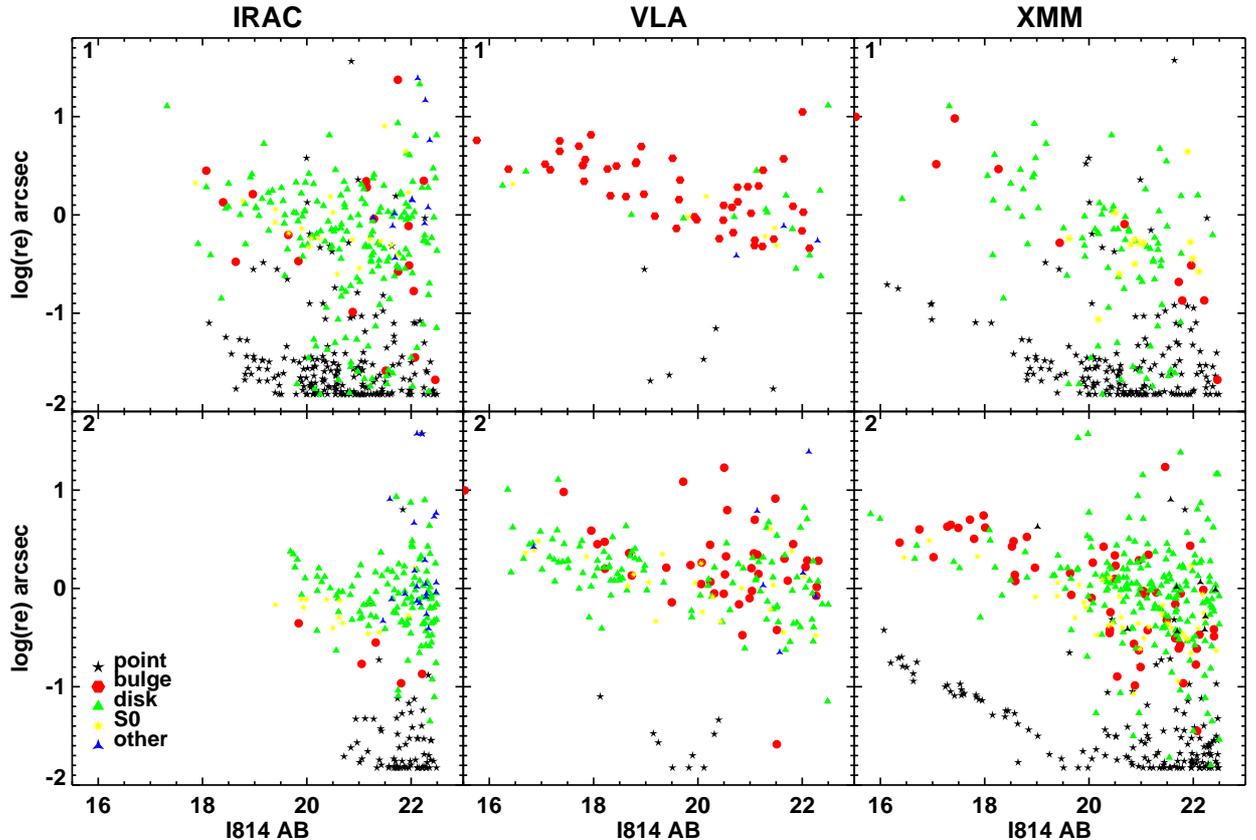}
\end{tabular}
\end{center}
\caption[CAPTION]{\label{fig:5} Half-light radius, $r_e$, versus
{\tt SExtractor} $I814$ MAG\_BEST as a function of visual morphological
classification.}
\end{figure*}

Figure~4 shows ten representative color images of classified galaxy
morphologies. It also motivates the need for visual morphologies
of these types of galaxies. We can see that seven out of ten of
these sources would have been classified as bulge-dominated early-type
galaxies from S\'ersic indices alone (e.g., $n > 2.5$).  Two of
these sources would have been classified as late-type galaxies or
disk galaxies, one of which is a point source with an unreliable
S\'ersic index. The other point source would have been classified
as bulge+disk.  The S0 galaxies are usually included in the $n >
2.5$, bulge-dominated sample, while the visual inspection allows
this unique class to be robustly identified.  For galaxies not
dominated by a bright, compact nucleus, the quantitative morphologies
generally agree with the visual morphologies.  Figure~5 plots
magnitude versus size as a function of visual type and AGN subsample
(similar to Fig.~2, except coded by visual morphology rather than
CLASS\_STAR).  We present the results of the visual morphological
classifications in Table 5.

Visually classifying these galaxies allows us to split the sample
into more discrete morphological bins rather than the three offered
by the S\'ersic index. For example, sources which were classified
as bulge-dominated, early-type galaxies due to a high S\'ersic index
are often classified as disk galaxies visually. From Table~5 we can
see that IRAC1 and XMM1 have high fractions of point sources (47\%
and 61\%, respectively), low fractions of bulge/early-type galaxies
(4\% for both), and large fractions of disk-dominated galaxies (41\%
and 31\%, respectively). IRAC2 and XMM2 have slightly lower fractions
of point sources (30\% and 31\%, respectively) and larger fractions
of disk galaxies (53\% and 46\%, respectively).  IRAC2 has very few
(2\%) bulge/early-type galaxies, while aside from the VLA samples,
XMM2 has the largest fraction (12\%) of bulge/early-type galaxies.
VLA1 has the largest fraction (59\%) of bulge/early-type galaxies.
This sample also has a large fraction (22\%) of disk galaxies.  VLA2
is dominated by disk galaxies (57\%) and 22\% of these sources are
classified as bulge/early-type galaxies.  The S0 and other
classification contribute small percentages to all of the samples.

% TABLE 5
\begin{deluxetable}{ccccccc}
\tablewidth{0pt}
\tablecaption{Visual AGN morphologies ($I814 \le 22.5$).}
\tablehead{
\colhead{Sample} &
\colhead{IRAC1} &  
\colhead{IRAC2} &
\colhead{XMM1} &
\colhead{XMM2} &
\colhead{VLA1} &
\colhead{VLA2}}
\startdata
$N_{\rm tot}$    & 444        & 238        & 280        & 458        & 81        & 188        \\
point source     & 209 (47\%) &  72 (30\%) & 175 (61\%) & 147 (31\%) &  6  (7\%) &   9  (5\%) \\
bulge/early-type &  19  (4\%) &   5  (2\%) &  11  (4\%) &  55 (12\%) & 48 (59\%) &  42 (22\%) \\
disk/late-type   & 184 (41\%) & 127 (53\%) &  86 (31\%) & 211 (46\%) & 18 (22\%) & 108 (57\%) \\
S0               &  21  (5\%) &  13  (6\%) &  13  (5\%) &  39  (9\%) &  6  (7\%) &  22 (12\%) \\
other            &  11  (3\%) &  21  (9\%) &   0  (0\%) &   9  (2\%) &  3  (4\%) &   7  (4\%) \\
\enddata
\end{deluxetable}

Comparing the results from the visual classifications (Table~5) to
the quantitative classifications (Table~4, for the $I814 \leq 22.5$
sample) shows a large discrepancy.  We see that the point sources
are identified equally well using either quantitative morphologies
or visual classifications, while the late-type and early-type
classifications often differ considerably.  Quantitatively, we find
an overabundance of bulge-dominated galaxies while our visual
classification shows that a large fraction of the galaxies are
actually disk galaxies.  AGN host galaxies comprise a very small
fraction of the total galaxy sample, but these results indicate
that more complex methods in analyzing such galaxies are required.

\section{Interesting Sources}

While visually inspecting the COSMOS AGN candidates for this analysis
--- both the magnitude-limited inspections in \S 4.2 as well as
considering outliers in \S 4.1 --- we identified a number of peculiar
or rare AGN candidates, discussed here.  We find a small class of
AGN hosted by face-on disk galaxies with compact, bright nuclear
sources (\S 5.1).  We find several galaxies which appear to host
two or more nuclear point sources (\S 5.2).  Such sources are likely
candidates for the late stages of the merger of two SMBHs, such as
the COSMOS dual AGN confirmed in \citet[][see also Civano et al.
2010]{com09}.  We identify several AGN host galaxies which appear
to contain an offset bright point source (\S 5.3).  These sources
may represent earlier phases of SMBH mergers, but might also be
supernovae or chance superpositions.  We also identify a handful
of lensed AGN candidates, or otherwise interesting sources (\S 5.4).

\subsection{Disk galaxies with a compact nucleus}

% FIGURE 6
\begin{figure*}[!t]
\begin{center}
\begin{tabular}{c}
\includegraphics[scale=0.5]{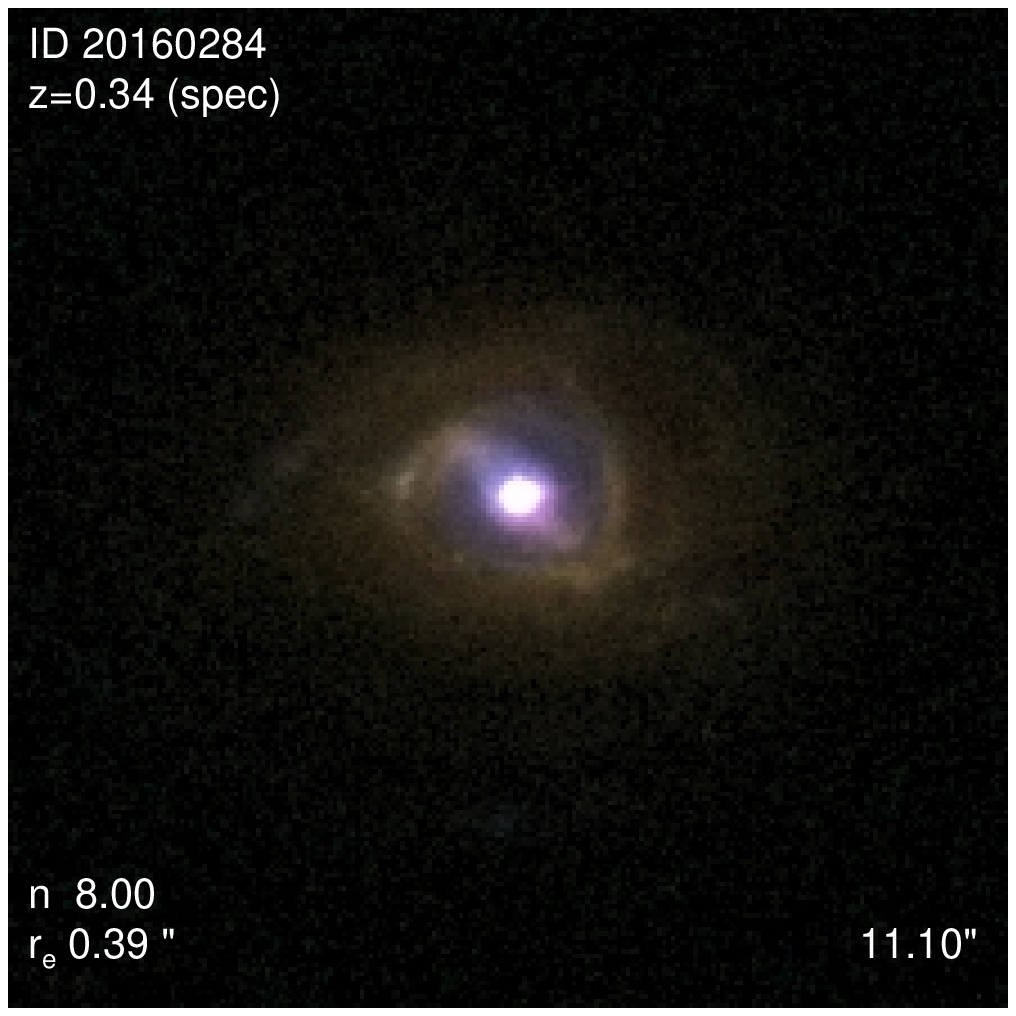}
\includegraphics[scale=0.5]{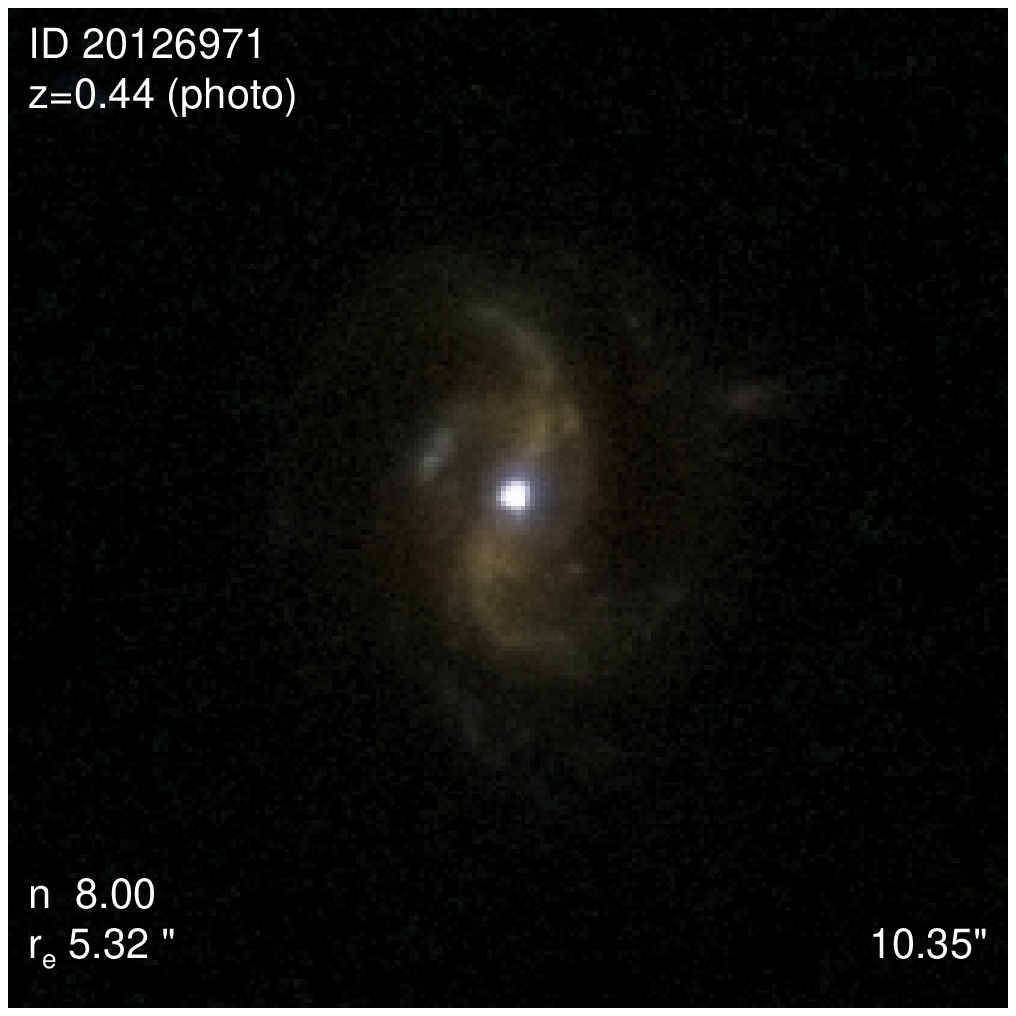}
\includegraphics[scale=0.5]{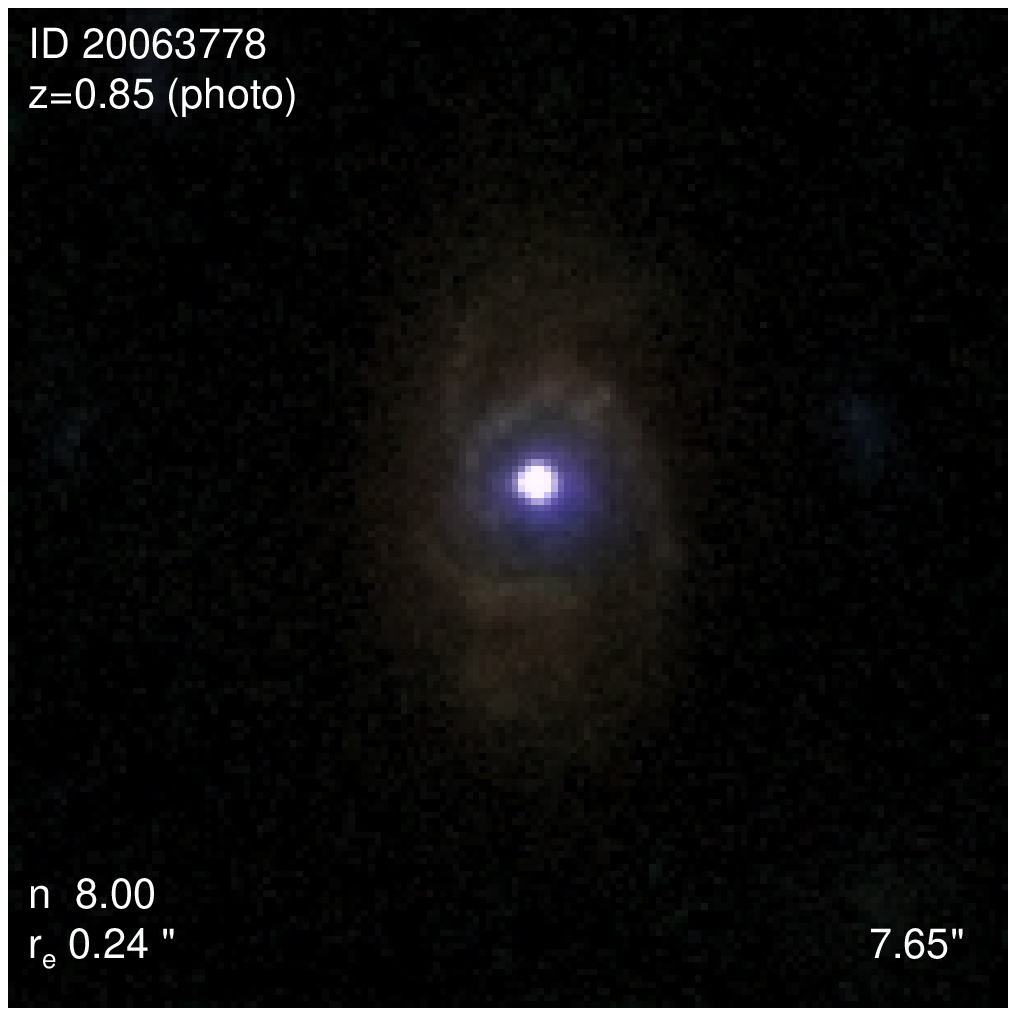}
\end{tabular}
\end{center}
\caption[CAPTION]{\label{fig:6} Three face-on spiral galaxies which
host an unobscured AGN. We visually identify just 41 such systems
(to $I814 = 22.5$).  These galaxies all have $n=8.0$ and extended
sizes.  ID~20160284 is one of the rare AGN identified by all three
selection methods.  Labels are as in Fig.~4.}
\end{figure*}

We visually identify 41 galaxies which appear to host a stong,
compact nucleus, suggestive of an unobscured AGN (see Table~6).  We
find 34/27/4 such systems identified using X-ray/IRAC/radio criteria,
respectively, with several sources identified using multiple criteria.
In particular, 20 of these sources were identified from both IRAC
and X-ray criteria and one source was identified using all three
selection criteria (ID~20160284; Fig.~6 left).  This last source
was classified as a quasar by SDSS, and shows broad H$\beta$ and
H$\alpha$ emission lines as well as narrow [\ion{O}{3}] emission
lines.  These galaxies generally represent a population of galaxies
known as Seyfert galaxies, presumably type~1 Seyferts based on the
strong, nuclear cores; however, spectra are required to distinguish
them as type~1 or type~2.  The rarity of these systems implies that
the unobscured AGN phase in most galaxies happens within a very
rapid timescale, e.g., the theoretical model of \cite{hop06} suggests
that the observable lifetime for a bright optical quasar is $\sim$
$10^7$ years, while the Seyfert-like galaxies identified here appear
to have lower nuclear luminosities than bright quasars.

% TABLE 6
\begin{deluxetable}{ccccccc}
\tablewidth{0pt}
\tablecaption{Disk galaxies with an unobscured AGN.}
\tablehead{
\colhead{ID} &
\colhead{RA} &
\colhead{Dec} &
\colhead{$z_{\rm spec}$} &
\colhead{$z_{\rm phot}$} &
\colhead{$I814$} &
\colhead{AGN Type}}
\startdata
  20001881 &  150.1731933  &  1.6163286  & \nodata & 0.57  & 19.51 & 4 \\
  20018586 &  149.9741093  &  1.6434971  &   1.03  & 1.11  & 20.78 & 9 \\
  20029957 &  150.5394505  &  1.9236122  & \nodata & 0.57  & 19.63 & 1 \\
  20033437 &  150.3272416  &  1.9285469  & \nodata & 0.57  & 20.19 & 5 \\
  20038068 &  150.1391437  &  1.8769805  & \nodata & 1.41  & 20.39 & 1 \\
  20040478 &  150.0249868  &  1.9147607  & \nodata & 0.94  & 20.65 & 9 \\
\enddata
\tablecomments{Coordinates are J2000.  $I814$ is the {\tt SExtractor}
MAG\_BEST magnitude (AB system).  AGN Type indicates which sample(s)
each source comes from: +1 - IRAC1, +2 - IRAC2, +4 - XMM1, +8 -
XMM2, +16 - VLA1, and +32 - VLA2.  {\it The full list of 41 sources
is provided as an on-line table.}}
\end{deluxetable}

\subsection{Dual/multiple nuclei AGN}

% FIGURE 7
\begin{figure*}[!t]
\begin{center}
\begin{tabular}{c}
\includegraphics[scale=0.5]{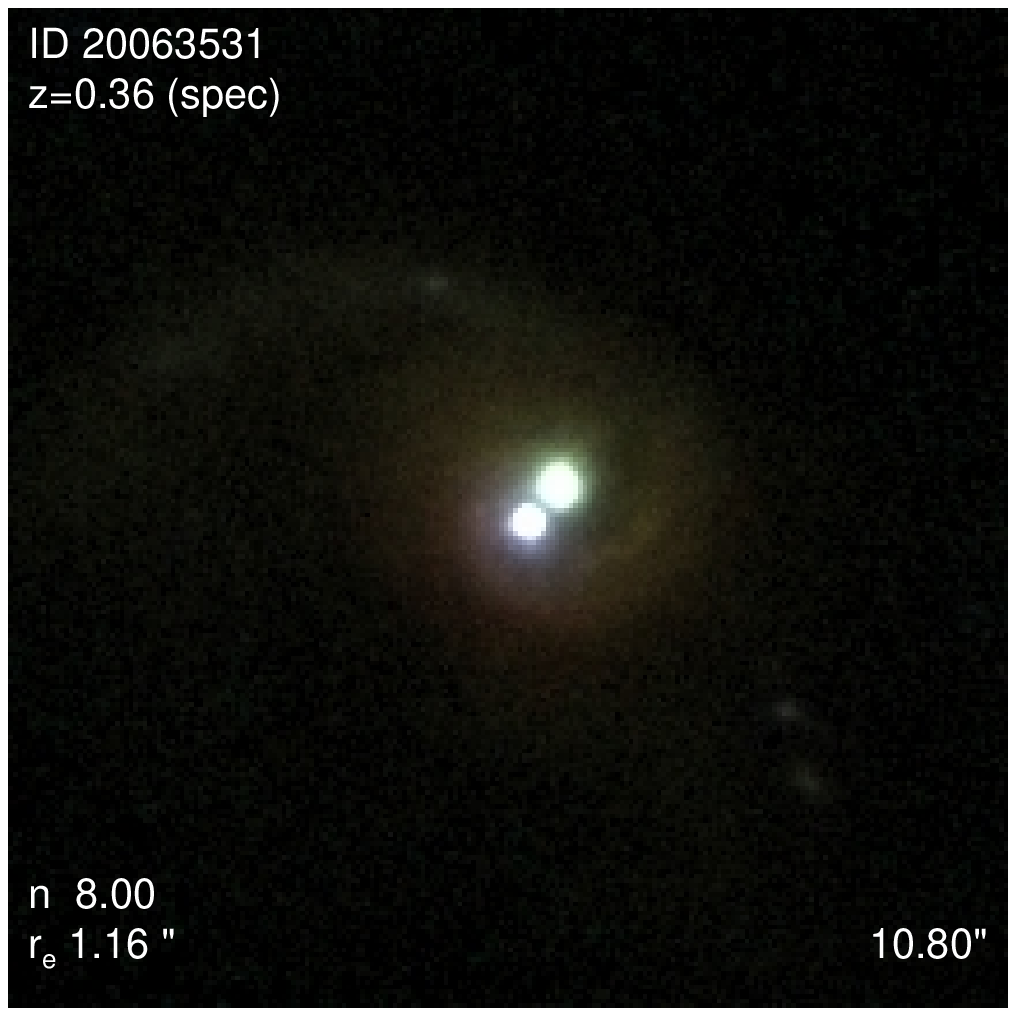}
\includegraphics[scale=0.5]{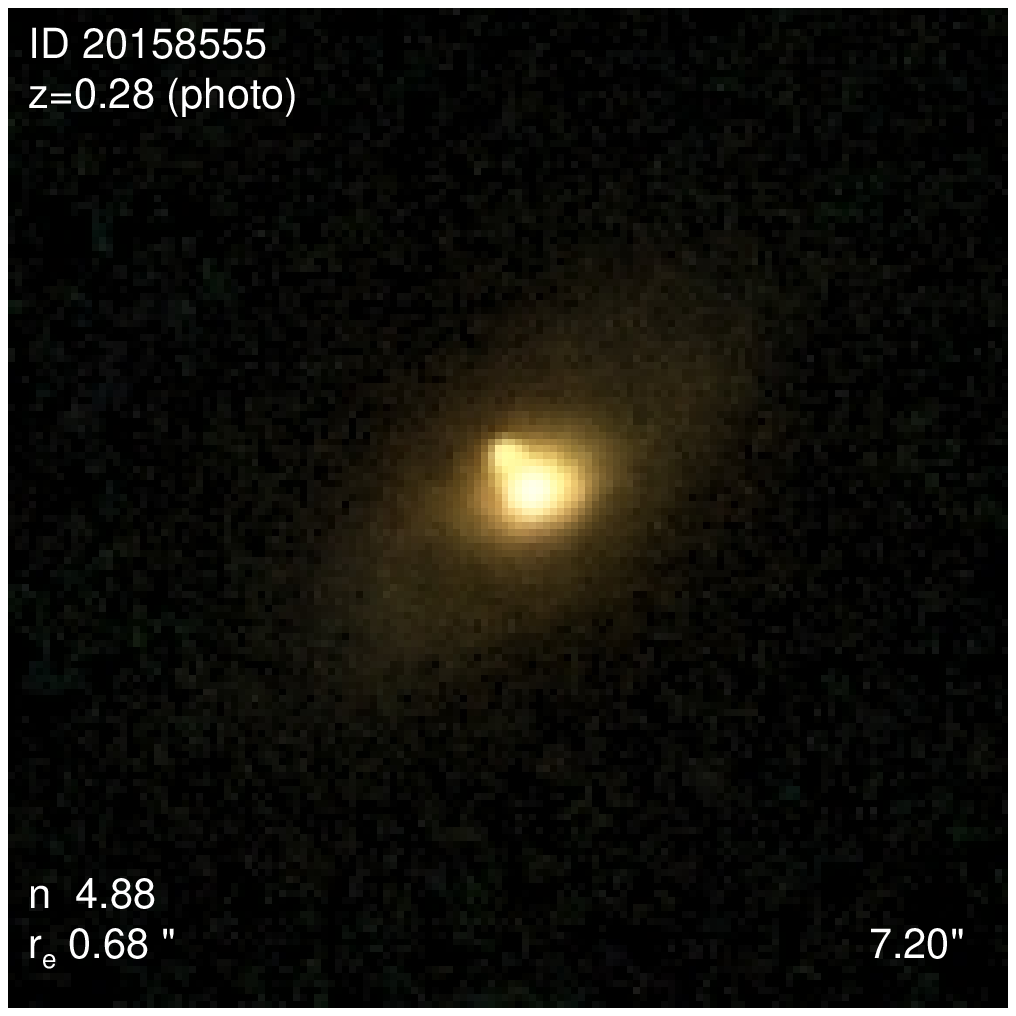}
\includegraphics[scale=0.5]{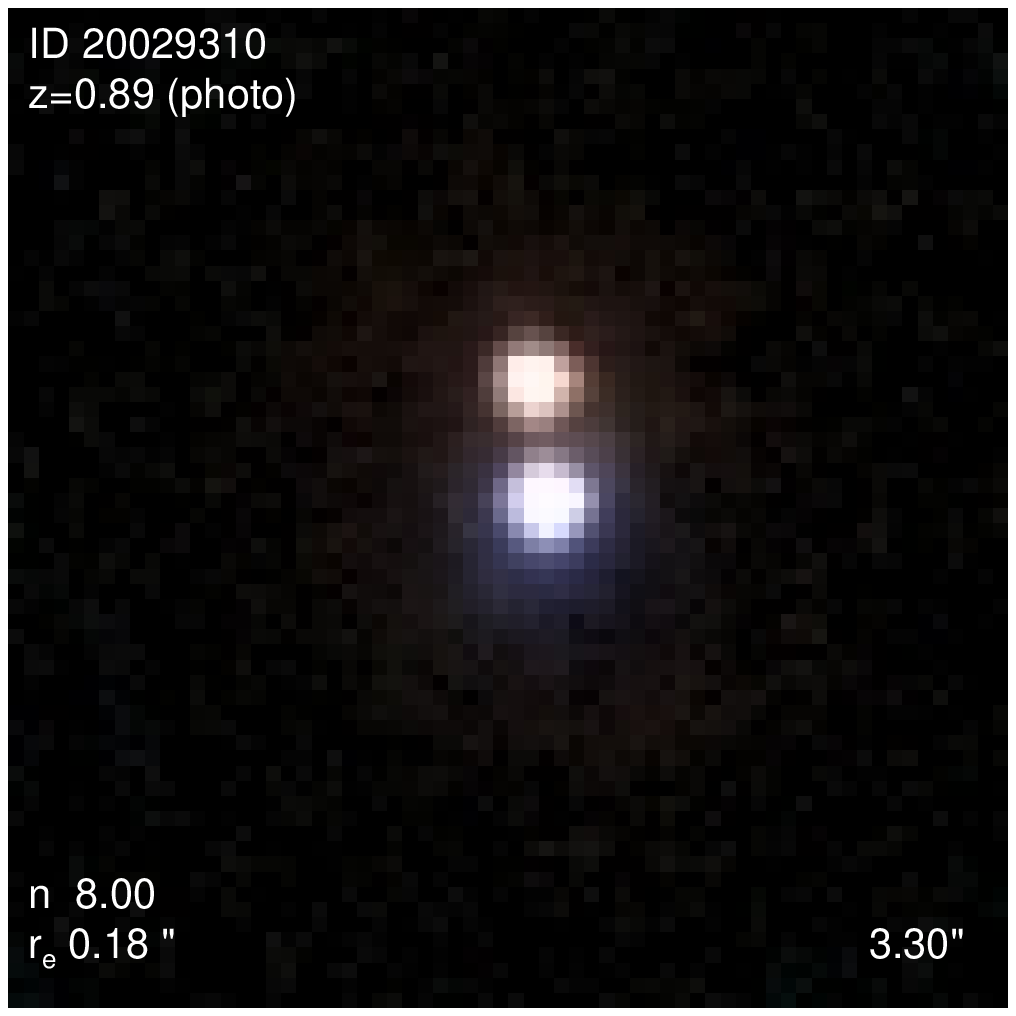}
\end{tabular}
\end{center}
\caption[CAPTION]{\label{fig:7} One confirmed and two candidate
dual AGN systems. Object~20063531 (COSMOS~J100043.15+020637.2) was
confirmed as a dual AGN in \cite{com09}.  Labels are as in Fig.~4.}
\end{figure*}

According to hierarchical galaxy formation theory, massive galaxies
assemble through the merging of smaller galactic components, with
smaller galaxies forming first and then merging into successively
larger bodies. It is also believed that SMBHs appear to reside at
the centers of most massive galaxies \citep{kor01}.  Given this
picture, we should often find two or more SMBHs in the process of
merging.  These merging SMBHs will initially be widely separated
($\ge$1 kpc) ``dual'' SMBHs.  After $\sim$ 100 Myr they will become
true binary SMBHs, gravitationally bound to one another with
parsec-scale separations; the merging into a single central SMBH
happens over much longer timescales \citep{beg80} and is expected
to be a major source of gravitational waves.  \cite{hop06} find
that the merging of two equal mass, gas rich systems can produce
or accentuate AGN or quasar activity. Though these systems should
be commonly observed, only a few cases have been discovered to date.

The classic example of a nearby dual AGN is NGC~6240 \citep{kom03}
which has two SMBHS spatially resolved by {\it Chandra}.  To date,
the most common method for detecting and identifying dual AGN has
been spectroscopically, in which multiple high-ionization state,
narrow emission lines are observed in a single AGN \citep[e.g.,][]{ger07,
com09a, xu09}.

Another method for finding these systems is with the use of high
resolution imaging.  By spatially resolving multiple nuclei within
a single galaxy, this method has the ability to conclusively
demonstrate that multiple emission lines in a spectrum are not due
to complex kinematics within a single AGN.  However, spectroscopy
is also required in order to demonstrate that the multiple nuclei
are not chance superpositions and that both are, indeed, active.
\cite{com09} reports the discovery of a dual AGN galaxy in the
COSMOS field which shows two bright point sources residing at the
center of a merger remnant spiral galaxy (J100043.15+020637.2;
Fig.~7, left).  The physical separation between the nuclei is
0.5\arcsec, or 2.5 kpc at the redshift of the galaxy, $z = 0.36$.
In addition to the {\it Hubble} imaging, observations from Keck,
zCOSMOS, SDSS, {\it XMM-Newton}, {\it Spitzer}, and the VLA fortify
the evidence for AGN activity in this galaxy \citep[see also][]{Civano:10}.
The discovery of this source has motivated us to visually search
and identify other potential candidates for dual AGN activity, and
we present three such candidates in Figure~7.  See Madsen et al.
(in prep.) for Keck spectroscopic observations of these candidates.

\subsection{Offset AGN candidates}

% FIGURE 8
\begin{figure*}[!t]
\begin{center}
\begin{tabular}{c}
\includegraphics[scale=0.5]{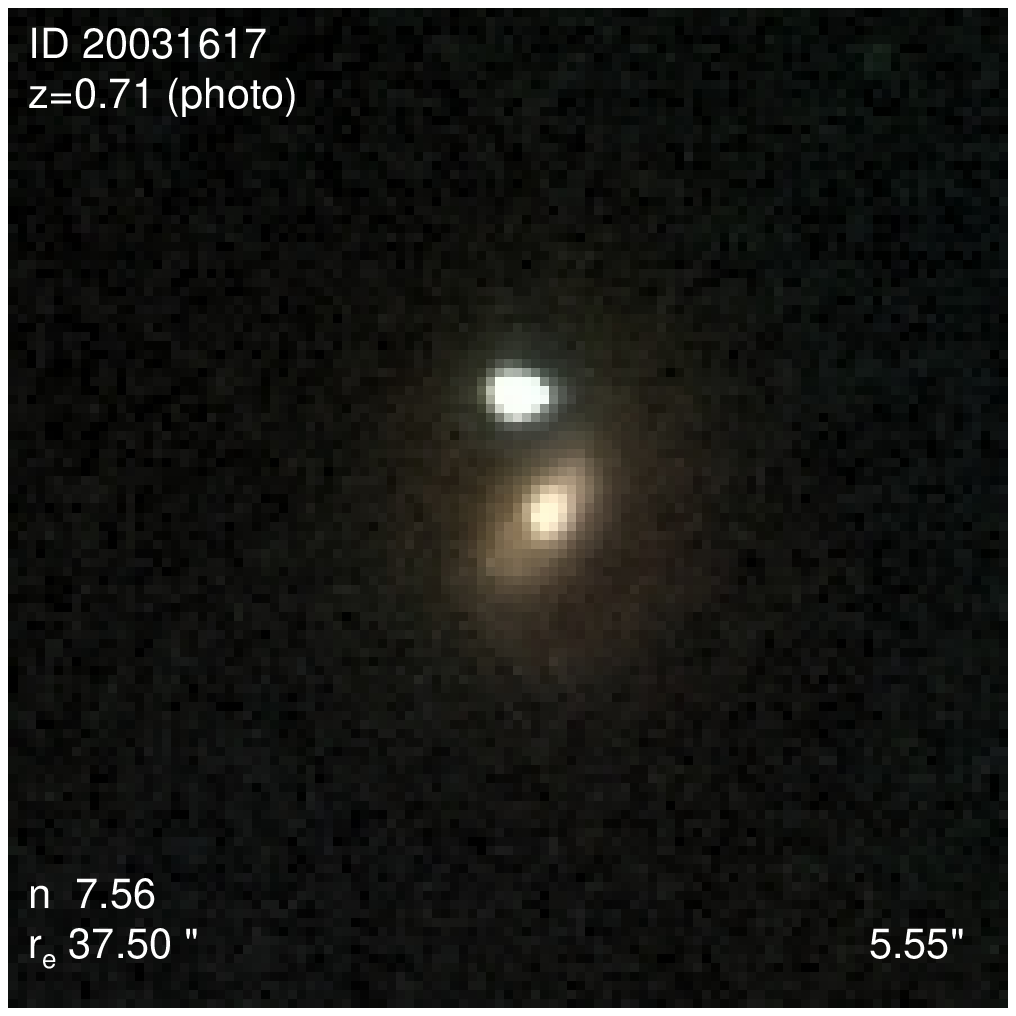}
\includegraphics[scale=0.5]{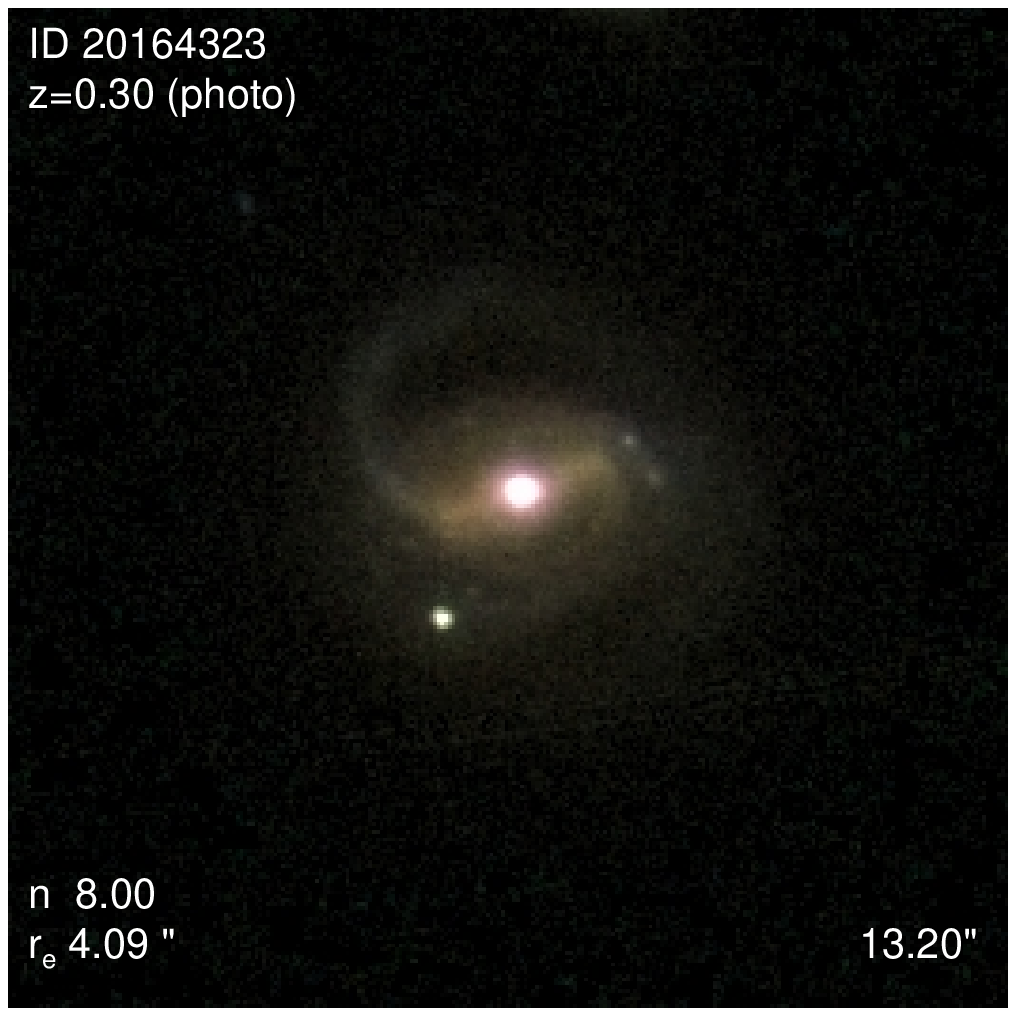}
\includegraphics[scale=0.5]{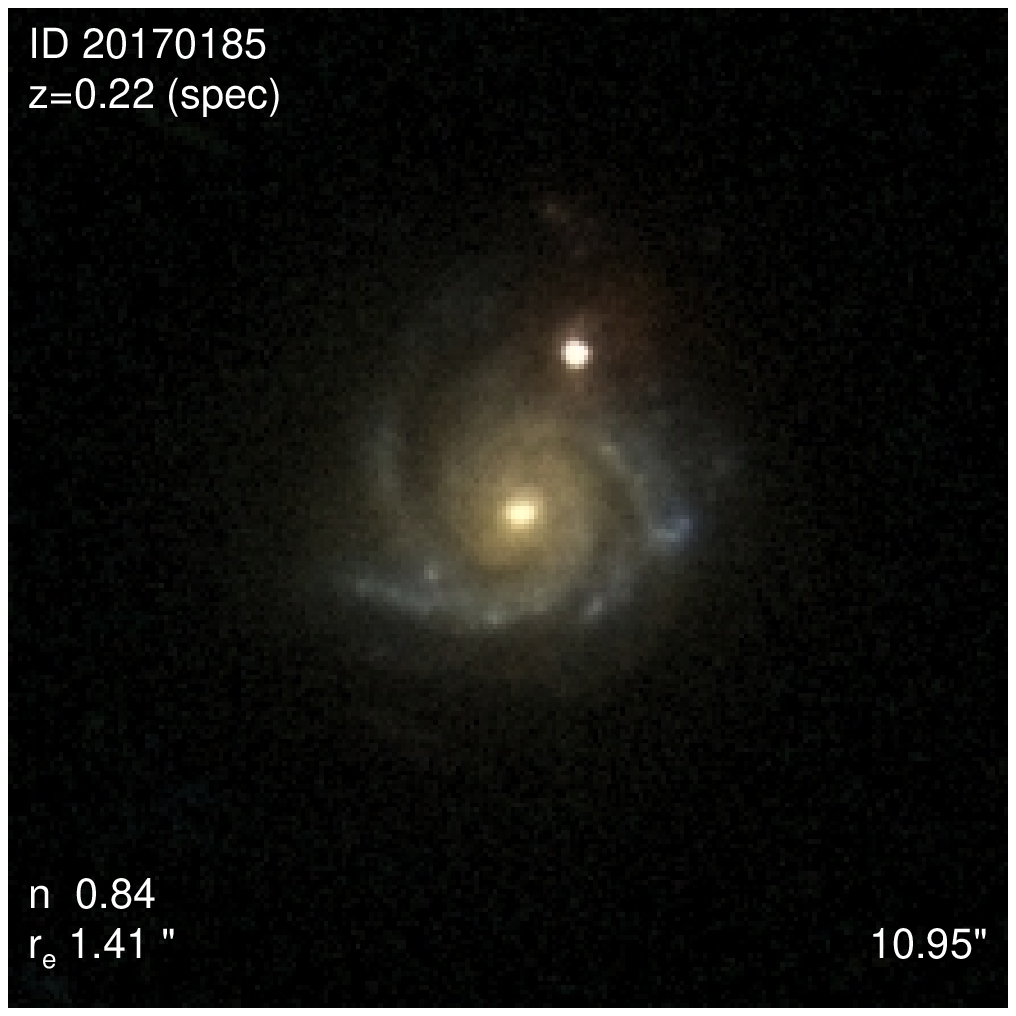}
\end{tabular}
\end{center}
\caption[CAPTION]{\label{fig:8} Three galaxies which potentially
host an offset AGN, though the offset source might also be either
a supernova or due to line of sight projections.  Spectroscopy
and/or high-resolution, multi-epoch imaging is required to determine
the nature of these systems.  Labels are as in Fig.~4.}
\end{figure*}

AGNs are not required to reside at the centers of galaxies and some
can be found offset from the galaxy center. \cite{bar08} report on
the discovery of a triplet of emission-line nuclei in the disturbed
galaxy NGC~3341. It is expected that minor mergers can also trigger
episodes of nuclear activity \citep{her95} with multiple AGN present
in some minor mergers so long as the secondary galaxies also harbor
SMBHs.  NGC~3341 appears to contain two dwarf galaxies merging with
a massive disk galaxy. One of the dwarf galaxies has a Seyfert~2
spectrum, while the other dwarf galaxy appears to have a LINER
spectrum. The primary nucleus appears to indicate low-level AGN
activity, making this a good candidate for a dual, or even a triple,
AGN.  Figure~8 shows three examples of galaxies which host an offset
source.  The bright, offset sources (often unresolved) could be
explained in one of three ways: (i) a supernova explosion in a
spiral arm, (ii) chance superposition, or (iii) an offset AGN,
potentially triggered by a minor merger.  Spectroscopy and/or
multi-epoch, high-resolution imaging is required to distinguish
between these interpretations.

\subsection{AGN lens candidates}

% FIGURE 9
\begin{figure*}[!t]
\begin{center}
\begin{tabular}{c}
\includegraphics[scale=0.5]{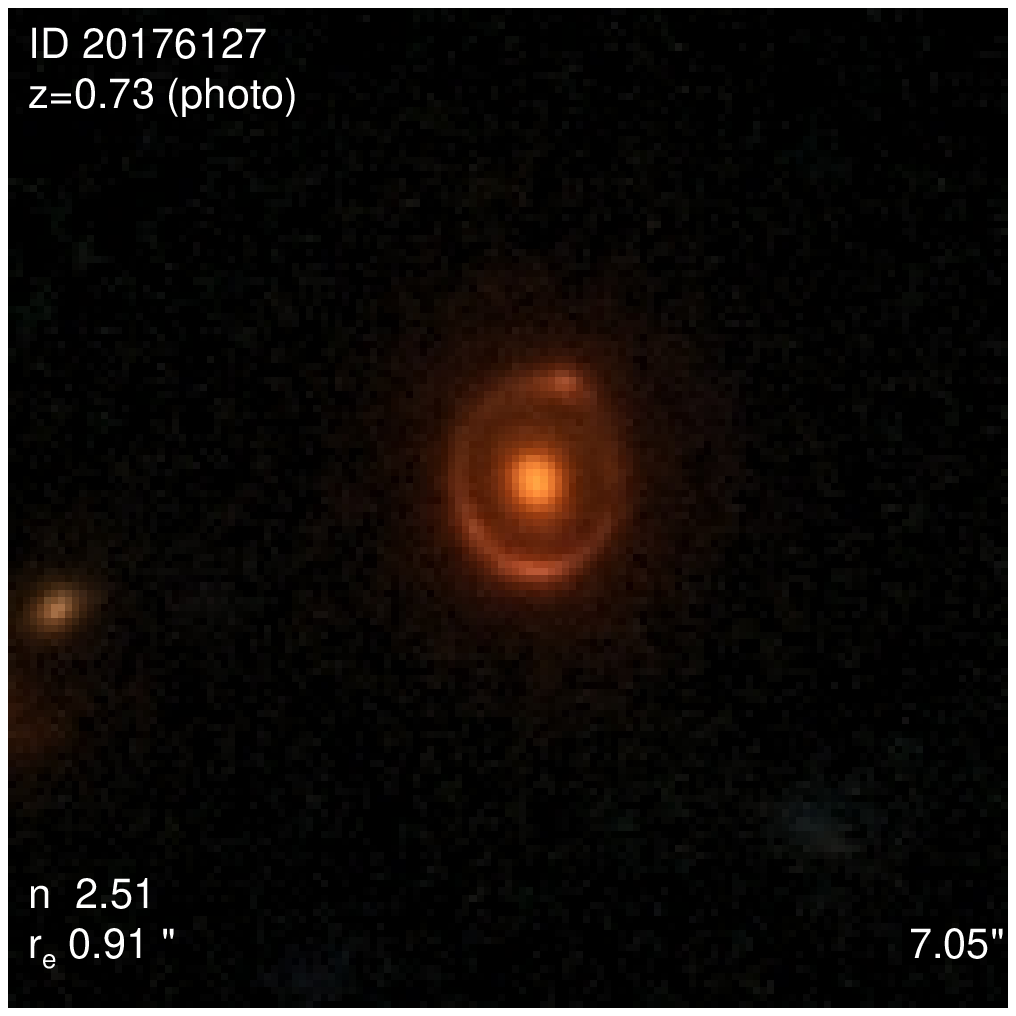}
\includegraphics[scale=0.5]{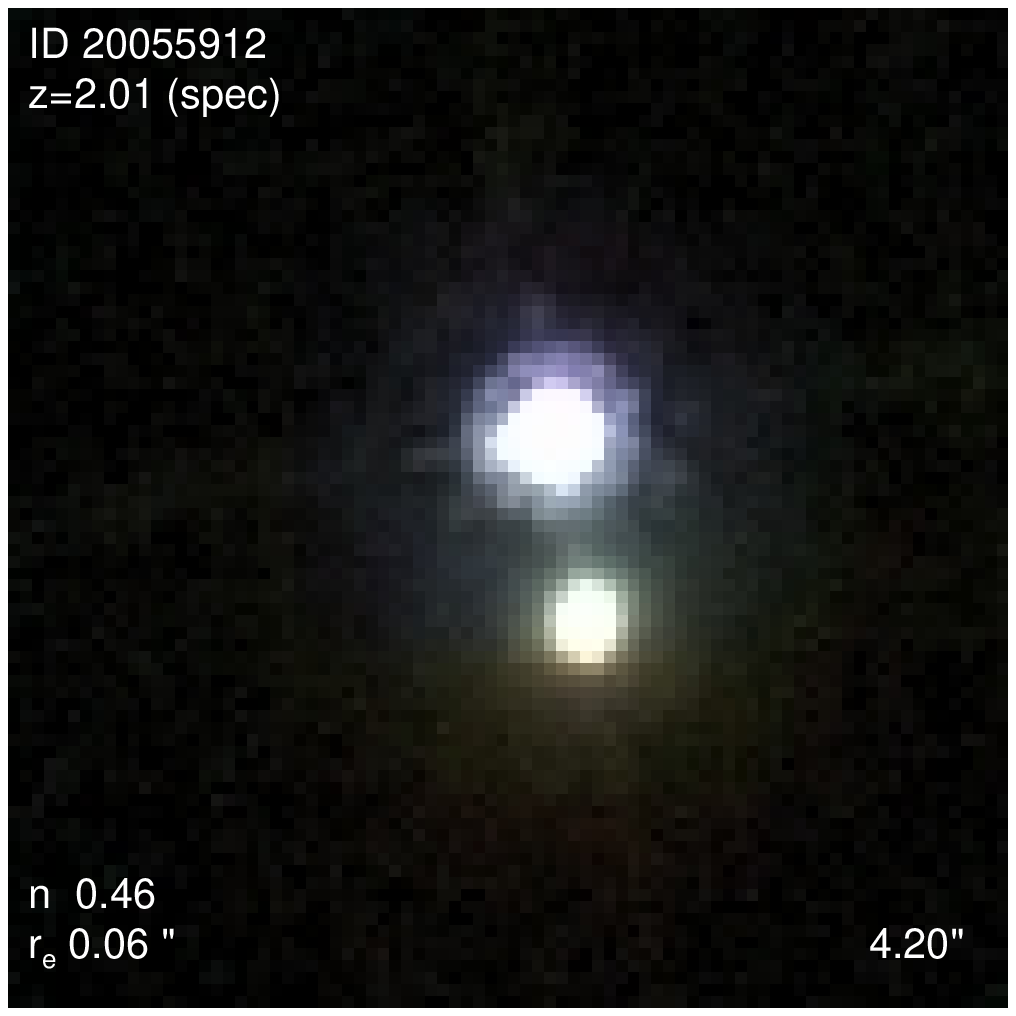}
\includegraphics[scale=0.5]{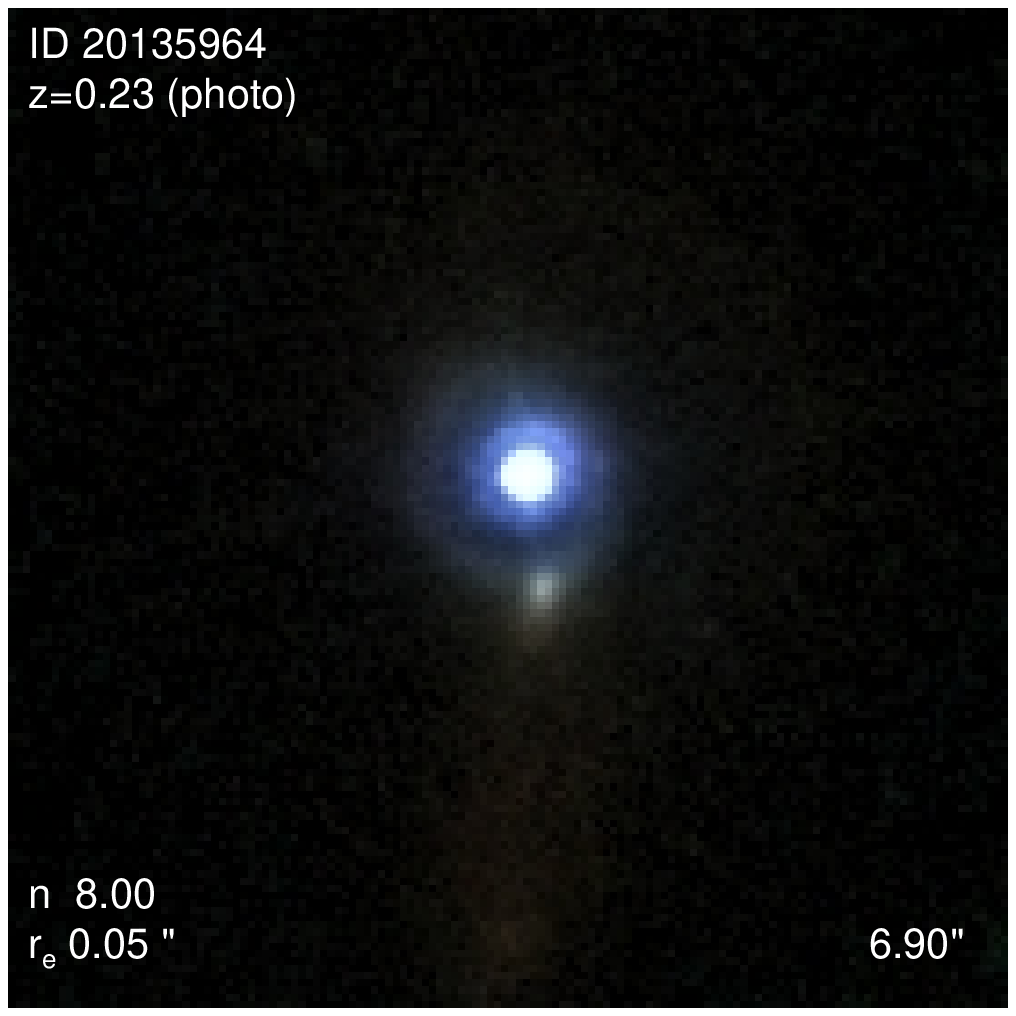}
\end{tabular}
\end{center}
\caption[CAPTION]{\label{fig:9} Three peculiar AGN identified in
this analysis, potentially representative of gravitational lensing.
Object~20176127 (left) appears to be an Einstein ring.  Note that
the similarly red colors of both the putative lensing source and
ring is likely due to the algorithm used to make these images,
painting colors at the resolution of the ground-based images onto
the high-resolution {\it Hubble} images (see \S 4.2).  Object~20055912
(middle) was identified as one of the 17 reddest quasars in the
SDSS survey in \citet{you08}.  These images show that the colors
are likely due to source blending rather than processes intrinsic
to the AGN.  Object~20135964 (right) is a peculiar AGN, with a
ring-like structure circling a blue point source.  Labels are as
in Fig.~4.}
\end{figure*}

We also find three AGN with peculiar morphologies suggestive of
gravitational lensing.  Lensed systems always provide the potential
for measuring the mass of the lensing galaxy.  However, since AGN
typically are time variable systems, rare lensed AGN systems provide
an additional cosmological probe through time delays
\citep[e.g.,][]{Schechter:97}.  COSMOS~100038.28+024133.8 (object
20176127; Fig.~9, left), which is a weak radio source, appears to
be a 1.2\arcsec\ diameter Einstein ring.  In fact, the source was
not included in the VLA AGN candidate sample (\S 3.1) since its
radio flux density is below 0.3~mJy and thus is potentially due to
processes related to star formation rather than nuclear activity.
This source was also identified in \citet{fau08} as one of 67
galaxy-galaxy strong lens candidates in the COSMOS field.

Cross-correlating the {\it XMM-Newton} archive with the SDSS third
data release (DR3) quasar catalog \citep{sch05}, \citet{you08}
identified SDSS~J100201.50+020329.4 (object 20055912; Fig.~9, middle)
as one of only 17 moderate-redshift ($0.9 < z < 2.1$) quasars with
extremely red optical colors ($g - r \geq 0.5$).  They attempt to
use the optical and X-ray data to discern whether the red colors
are caused by an unusual intrinsic continuum or by dust reddening.
Based on the moderately flat X-ray spectrum of SDSS~J1002+0203
($\Gamma \sim 1.9$) but relatively low level of dust reddening
inferred from the optical spectroscopy, \citet{you08} classify this
source as having an intrinsically red AGN continuum.  They further
suggest that the different physics of AGN with low accretion rates
might be responsible for the unusually red color.  The low spatial
resolution of the SDSS imaging, however, misses the nearby companion,
offset by 0.8\arcsec\ to the SSW, which is clearly visible in the
COSMOS {\it Hubble} imaging.  This presumably foreground galaxy
undoubtedly affects the SDSS photometry of the quasar.  However,
assuming that the second object is a foreground early-type galaxy
with minimal gas or dust, it would not affect the X-ray spectrum
of the quasar.  The foreground galaxy would also presumably be
lensing the background quasar at some level (though not strongly,
since no counterimages are evident).  It would be interesting to
obtain high-resolution images of other quasars in the \citet{you08}
sample to see if they could also be explained by chance superpositions
and source blending in the SDSS photometry rather than unusual AGN
continua.

Finally, COSMOS~100232.13+023537.3 (object 20135964; Fig.~9, right)
presents a very unusual morphology with a blue unresolved core
surrounded by a 1.4\arcsec\ diameter ring-like structure with a
southern extension.  This AGN candidate is one of only 23 (out of
2413) AGN candidates selected independently by all three selection
criteria.  The structure is initially suggestive of a gravitational
lens, albeit with the atypical configuration of the quasar being
the lens, not the lensed background source \citep[e.g.,][]{Courbin:10}.
However, a recent spectrum obtained with the Low Resolution Imaging
Spectrograph \citep[LRIS;][]{Oke:95} on the Keck~I telescope reveals
that both the quasar and the southern knot are at the same redshift,
implying that the unusual morphology is due to merger activity and
not gravitional lensing.

\section{Discussion}

In this paper, we use the ACS-GC data set in combination with radio,
X-ray, and mid-IR imaging in the COSMOS field to investigate the
morphological diversity of AGN host galaxies.  We utilize high
resolution images obtained with the {\it Hubble} ACS instrument to
both quantitatively ({\tt GALFIT}) and qualitatively (visually, to
$I814 = 22.5$) segregate our AGN sample into discrete morphological
bins.  We use the radio, X-ray, and mid-IR imaging to select our
AGN candidates and separate the samples into six categories, two
per wavelength and split by flux density in the AGN selection band.

We investigate the overlap between the AGN samples and find significant
overlap for the unresolved AGN candidates.  The majority of the
unresolved radio AGN are also selected by either X-ray emission or
mid-IR colors, and there is also considerable overlap between the
X-ray and mid-IR unresolved sources.  However, only a small fraction
($2 - 5\%$) of the unresolved AGN selected in the X-rays or mid-IR
are selected as radio AGN, consistent with a minority fraction of
AGN being radio loud.  Only a small fraction of the optically
resolved AGN samples are identified using multiple techniques.

The radio-selected AGN are the most distinct population, with a
relatively low incidence of unresolved optical morphologies, e.g.,
classical quasars ($5 - 9\%$).  In contrast, approximately half of
the bright X-ray and mid-IR selected AGN samples are optically
unresolved (ranging from 44\% to 61\%), and nearly a third of the
fainter X-ray and mid-IR selected AGN samples are optically unresolved
(ranging from 19\% to 33\%).  The broad band spectral energy
distributions of quasars are typically dominated by nuclear emission,
and such systems are generally believed to be emitting near their
Eddington limit \citep[e.g.,][]{kol06}.  This is consistent with
the correlations between optical brightness and X-ray and mid-IR
brightness that we observe in Fig.~3 for the X-ray and mid-IR
samples.  In contrast, the radio-selected AGN show no correlation
between their optical and radio brightnesses, suggesting that such
sources represent AGN accreting at a range of Eddington ratios and
experience SMBH accretion mechanisms distinct from the quasars.

From our visual classifications, we find that the brighter
radio-selected AGN generally (59\%) reside in bulge-dominated,
early-type galaxies, consistent with a literature that stretches
back nearly 50 years \citep{Matthews:64, Pentericci:01, Best:05}.
Fainter radio sources, however, preferentially (57\%) reside in
disk galaxies.  Note, however, that the two highest fractions of
bulge-dominated hosts are consistently the two radio samples.  The
X-ray and mid-IR AGN samples are most likely to be associated with
point sources or disk galaxies, with such identifications accounting
for between 77\% and 92\% of the samples.  For the brighter X-ray
and mid-IR samples, point source identifications are more common,
while the fainter X-ray and mid-IR samples are more likely to be
hosted by disk galaxies.  The fraction of these samples associated
with early-type or S0 galaxies is low, accounting for $8 - 9\%$ of
the mid-IR selected samples and a similar fraction for the bright
X-ray sample.  In contrast, the fainter X-ray sample has a much
higher rate of being hosted by early-type or S0 galaxies, 21\%.

These morphological results are consistent with \cite{hic09}, who
studied the colors, luminosities, and clustering of AGN selected
by different criteria.  \citet{hic09} find radio-selected AGN tend
to reside in luminous red sequence galaxies, are strongly clustered,
and have low Eddington ratios.  This is consistent with our result
that radio-selected AGN are the most likely to have elliptical or
early-type morphologies.  \citet{hic09} find X-ray selected AGN
reside in less clustered environments, have higher Eddington ratios
than radio AGN, and preferentially occupy the ``green valley'' of
color-magnitude space.  For mid-IR selected AGNs, \citet{hic09}
find them to be weakly clustered, have Eddington ratios similar to
X-ray AGN, and to reside in slightly bluer, less luminous galaxies
(e.g., reside in the ``blue cloud'').  Our morphological analysis,
considering samples selected over a wider redshift range and from
a deeper survey, support these results.  We find that the X-ray and
mid-IR AGN candidates broadly share similar morphologies, though
the X-ray AGN, particularly the fainter X-ray AGN, have a much
higher incidence of being hosted by early-type or S0 galaxies.  In
the general framework where AGN activity marks or regulates the
transition from late-type galaxies into massive, red sequence
galaxies, this work suggests a statistical chronological progression
from mid-IR selected AGN to X-ray selected AGN, both often accreting
at near Eddington rates, to eventual residence on the red sequence
where smaller bursts of nuclear activity are at non-Eddington rates
and associated with radio emission.

We have investigated the morphological properties of a large sample
of AGN, and analyzed how these morphologies depend on how the AGN
were selected.  A more detailed analysis attempting to more firmly
place these results within the context of galaxy formation and
evolution would require (near-) complete spectroscopic data for
both the AGN and their non-active counterparts, work that is actively
being done in the COSMOS field.  Spectroscopy would allow us to
confirm the nuclear activity through emission line diagnostics,
split the samples between unobscured and obscured AGN, as well as
investigate evolutionary trends.  We include the morphological
classifications of this large data set in the Appendix to support
that future work.

\acknowledgements

We thank the COSMOS team for their exhaustive work on and public
release of this important survey.  In particular, we thank Peter
Capak for useful discussions and for allowing us use of the pseudocolor
images derived from the Subaru plus {\it Hubble} images.  We also
thank Leonidas Moustakas for encouragement and useful discussions,
Jennifer Lotz for useful suggestions, and the anonymous referee for
a timely and helpful report.  Finally, we thank Fiona Harrison,
Mansi Kasiwal, and Shri Kulkarni for obtaining a spectrum of
ID~20135964.  The work of RLG and DS was carried out at the Jet
Propulsion Laboratory, California Institute of Technology, under a
contract with NASA.

\appendix
\section{Electronic Catalog of AGN Candidates}

We include two large electronic catalogs:  Table~7 provides the
morphological information for the full sample of 2413 AGN candidates
discussed in this paper.  Table~8 provides useful phomometric and
spectroscopic information for the same sources.

% TABLE 7
\begin{deluxetable}{ccccccccc}
\tablewidth{0pt}
\tablecaption{Morphological parameters for AGN candidates.}
\tablehead{
\colhead{ID} &
\colhead{RA} &
\colhead{Dec} &
\colhead{CLASS\_STAR} &
\colhead{$I814$} &
\colhead{$r_e$ (\arcsec)} &
\colhead{$n$} &
\colhead{VIS} &
\colhead{AGN Type}}
\startdata
20000085  & 150.7105669    & 1.6054066 &    0.00   & 20.19   &  0.64 &    0.24          &         B &       32\\
20000120  & 150.6741353    & 1.5972655    & 0.03   & 21.01    & 0.38   &  0.83          &         B     &    8\\
20000338  & 150.7121413    & 1.5881418    & 0.00   & 23.72    & 8.16   &  5.24        &  \nodata & 1\\
20000360  & 150.6680447    & 1.6164943    & 0.03   & 19.99    & 3.56   &  7.99      &             B       &  8\\
20000364  & 150.6854324   &  1.6159961    & 0.00   & 21.80   &  1.02   &  2.23    &               B       &  1\\
20000417  & 150.6238949 &    1.6036156    & 0.03   & 20.72 &    0.67   &  4.17 &                  B       &  8\\
\enddata
\tablecomments{Coordinates are J2000.  $I814$ is the {\tt SExtractor}
MAG\_BEST magnitude (AB system).  VIS lists the visual morphological
classification for sources brighter than $I814 = 22.5$:  A =
bulge-dominated, B = disk, C = unresolved (point source), D = S0,
and E = other.  AGN Type indicates which sample(s) each source comes
from: +1 - IRAC1, +2 - IRAC2, +4 - XMM1, +8 - XMM2, +16 - VLA1, and
+32 - VLA2.  {\it The full list of 2413 sources is provided as an
on-line table.}}

\end{deluxetable}

% TABLE 8
\begin{deluxetable}{ccccccccccc}
\tablewidth{0pt}
\tablecaption{Photometric and spectroscopic parameters for AGN candidates.}
\tablehead{
\colhead{ID} &
\colhead{[3.6]} &
\colhead{[4.5]} &
\colhead{[5.8]} &
\colhead{[8.0]} &
\colhead{$S_{0.5 - 2.0}$} &
\colhead{$S_{2 - 10}$} &
\colhead{$S_{5 - 10}$} &
\colhead{$S_{1.4}$} &
\colhead{$z_{\rm spec}$} &
\colhead{$z_{\rm phot}$}}
\startdata
20000085 & \nodata &  \nodata & \nodata & \nodata & \nodata & \nodata&  \nodata &    0.34 & \nodata &    0.62 \\
20000120 & \nodata & \nodata & \nodata & \nodata &   -0.15  &   3.37  &   2.44 & \nodata  &\nodata  &   0.65 \\
20000338 & 17.30   &  16.26   & 15.17  &  14.02 & \nodata & \nodata & \nodata & \nodata  &\nodata  &   1.96 \\
20000360 & \nodata & \nodata & \nodata&  \nodata &    0.19 &    0.96  &  -0.46 & \nodata   &  0.37  &   0.38 \\
20000364 & 16.34   & 15.87   & 15.38  &  14.35 & \nodata & \nodata & \nodata  &\nodata  &\nodata &    1.02 \\
20000417 & \nodata & \nodata & \nodata & \nodata &   -0.14 &    0.75 &    1.28 & \nodata & \nodata   &  0.98 \\
\enddata
\tablecomments{IRAC magnitudes are in the Vega system.  {\it
XMM-Newton} flux densities ($S_{0.5 - 2.0}$, $S_{2 - 10}$ and $S_{5
- 10}$, in the $0.5 - 2$~keV, $2 - 10$~keV and $5 - 10$~keV bands,
respectively) are in units of $10^{-14}~ {\rm erg}~ {\rm cm}^{-2}~
{\rm s}^{-1}$.  Radio flux densities at 1.4~GHz ($S_{1.4}$) are in
units of mJy.  {\it The full list of 2413 sources is provided as
an on-line table.}}

\end{deluxetable}

\bibliographystyle{apj.bst}
\bibliography{apj-jour,mybib}

\begin{thebibliography}{57}
\expandafter\ifx\csname natexlab\endcsname\relax\def\natexlab#1{#1}\fi

\bibitem[{{Alonso-Herrero} {et~al.}(2006)}]{alo06}
{Alonso-Herrero}, A. {et~al.} 2006, \apj, 640, 167

\bibitem[{{Ashby} {et~al.}(2009)}]{ash09}
{Ashby}, M.~L.~N. {et~al.} 2009, \apj, 701, 428

\bibitem[{{Barth} {et~al.}(2008){Barth}, {Bentz}, {Greene}, \& {Ho}}]{bar08}
{Barth}, A.~J., {Bentz}, M.~C., {Greene}, J.~E., \& {Ho}, L.~C. 2008, \apjl,
  683, L119

\bibitem[{{Becker} {et~al.}(1995){Becker}, {White}, \& {Helfand}}]{bec95}
{Becker}, R.~H., {White}, R.~L., \& {Helfand}, D.~J. 1995, \apj, 450, 559

\bibitem[{{Begelman} {et~al.}(1980){Begelman}, {Blandford}, \& {Rees}}]{beg80}
{Begelman}, M.~C., {Blandford}, R.~D., \& {Rees}, M.~J. 1980, \nat, 287, 307

\bibitem[{{Bertin} \& {Arnouts}(1996)}]{ber96}
{Bertin}, E. \& {Arnouts}, S. 1996, \aaps, 117, 393

\bibitem[{Best {et~al.}(2005)Best, Kauffmann, Heckman, Brinchmann, Charlot,
  Ivezi{\'c}, \& White}]{Best:05}
Best, P.~N., Kauffmann, G., Heckman, T.~M., Brinchmann, J., Charlot, S.,
  Ivezi{\'c}, {\'Z}., \& White, S.~D.~M. 2005, \mnras, 362, 25

\bibitem[{{Cappelluti} {et~al.}(2009)}]{cap09}
{Cappelluti}, N. {et~al.} 2009, \aap, 497, 635

\bibitem[{Civano {et~al.}(2010)}]{Civano:10}
Civano, F. {et~al.} 2010, \apj, submitted (astro-ph/1003.0020)

\bibitem[{{Comerford} {et~al.}(2009{\natexlab{a}}){Comerford}, {Griffith},
  {Gerke}, {Cooper}, {Newman}, {Davis}, \& {Stern}}]{com09}
{Comerford}, J.~M., {Griffith}, R.~L., {Gerke}, B.~F., {Cooper}, M.~C.,
  {Newman}, J.~A., {Davis}, M., \& {Stern}, D. 2009{\natexlab{a}}, \apjl, 702,
  L82

\bibitem[{{Comerford} {et~al.}(2009{\natexlab{b}})}]{com09a}
{Comerford}, J.~M. {et~al.} 2009{\natexlab{b}}, \apj, 698, 956

\bibitem[{Courbin {et~al.}(2010)Courbin, Tewes, Sluse, Meylan, Djorgovski,
  Mahabal, \& Rerat}]{Courbin:10}
Courbin, F., Tewes, M., Sluse, D., Meylan, G., Djorgovski, S.~G., Mahabal, A.,
  \& Rerat, F. 2010, \apj, submitted (astro-ph/1002.4991)

\bibitem[{{Donley} {et~al.}(2008){Donley}, {Rieke}, {P{\'e}rez-Gonz{\'a}lez},
  \& {Barro}}]{don08}
{Donley}, J.~L., {Rieke}, G.~H., {P{\'e}rez-Gonz{\'a}lez}, P.~G., \& {Barro},
  G. 2008, \apj, 687, 111

\bibitem[{{Donley} {et~al.}(2007){Donley}, {Rieke}, {P{\'e}rez-Gonz{\'a}lez},
  {Rigby}, \& {Alonso-Herrero}}]{donley07}
{Donley}, J.~L., {Rieke}, G.~H., {P{\'e}rez-Gonz{\'a}lez}, P.~G., {Rigby},
  J.~R., \& {Alonso-Herrero}, A. 2007, \apj, 660, 167

\bibitem[{{Eckart} {et~al.}(2010){Eckart}, {McGreer}, {Stern}, {Harrison}, \&
  {Helfand}}]{eck10}
{Eckart}, M.~E., {McGreer}, I.~D., {Stern}, D., {Harrison}, F.~A., \&
  {Helfand}, D.~J. 2010, \apj, 708, 584

\bibitem[{{Eckart} {et~al.}(2006){Eckart}, {Stern}, {Helfand}, {Harrison},
  {Mao}, \& {Yost}}]{eck06}
{Eckart}, M.~E., {Stern}, D., {Helfand}, D.~J., {Harrison}, F.~A., {Mao},
  P.~H., \& {Yost}, S.~A. 2006, \apjs, 165, 19

\bibitem[{{Eisenhardt} {et~al.}(2004)}]{eis04}
{Eisenhardt}, P.~R. {et~al.} 2004, \apjs, 154, 48

\bibitem[{{Faure} {et~al.}(2008)}]{fau08}
{Faure}, C. {et~al.} 2008, \apjs, 176, 19

\bibitem[{{Fazio} {et~al.}(2004)}]{faz04}
{Fazio}, G.~G. {et~al.} 2004, \apjs, 154, 10

\bibitem[{{Ferrarese} \& {Merritt}(2000)}]{fer00}
{Ferrarese}, L. \& {Merritt}, D. 2000, \apjl, 539, L9

\bibitem[{{Gabor} {et~al.}(2009)}]{gab09}
{Gabor}, J.~M. {et~al.} 2009, \apj, 691, 705

\bibitem[{{Gebhardt} {et~al.}(2000)}]{geb00}
{Gebhardt}, K. {et~al.} 2000, \apjl, 539, L13

\bibitem[{{Georgakakis} {et~al.}(2009)}]{geo09}
{Georgakakis}, A. {et~al.} 2009, \mnras, 397, 623

\bibitem[{{Gerke} {et~al.}(2007)}]{ger07}
{Gerke}, B.~F. {et~al.} 2007, \apjl, 660, L23

\bibitem[{Giavalisco {et~al.}(2004)}]{gia04}
Giavalisco, M. {et~al.} 2004, \apj, 600, L93

\bibitem[{{Hasinger} {et~al.}(2007)}]{has07}
{Hasinger}, G. {et~al.} 2007, \apjs, 172, 29

\bibitem[{{H{\"a}ussler} {et~al.}(2007)}]{hau07}
{H{\"a}ussler}, B. {et~al.} 2007, \apjs, 172, 615

\bibitem[{{Hernquist} \& {Mihos}(1995)}]{her95}
{Hernquist}, L. \& {Mihos}, J.~C. 1995, \apj, 448, 41

\bibitem[{{Hickox} {et~al.}(2009)}]{hic09}
{Hickox}, R.~C. {et~al.} 2009, \apj, 696, 891

\bibitem[{{Hopkins} {et~al.}(2006){Hopkins}, {Hernquist}, {Cox}, {Di Matteo},
  {Robertson}, \& {Springel}}]{hop06}
{Hopkins}, P.~F., {Hernquist}, L., {Cox}, T.~J., {Di Matteo}, T., {Robertson},
  B., \& {Springel}, V. 2006, \apjs, 163, 1

\bibitem[{{Kauffmann} {et~al.}(2003)}]{kau03}
{Kauffmann}, G. {et~al.} 2003, \mnras, 346, 1055

\bibitem[{{Koekemoer} {et~al.}(2007)}]{koek07}
{Koekemoer}, A.~M. {et~al.} 2007, \apjs, 172, 196

\bibitem[{{Kollmeier} {et~al.}(2006)}]{kol06}
{Kollmeier}, J. {et~al.} 2006, \apj, 648, 128

\bibitem[{{Komossa} {et~al.}(2003){Komossa}, {Burwitz}, {Hasinger}, {Predehl},
  {Kaastra}, \& {Ikebe}}]{kom03}
{Komossa}, S., {Burwitz}, V., {Hasinger}, G., {Predehl}, P., {Kaastra}, J.~S.,
  \& {Ikebe}, Y. 2003, \apjl, 582, L15

\bibitem[{{Kormendy} \& {Gebhardt}(2001)}]{kor01}
{Kormendy}, J. \& {Gebhardt}, K. 2001, in American Institute of Physics
  Conference Series, Vol. 586, 20th Texas Symposium on relativistic
  astrophysics, ed. {J.~C.~Wheeler \& H.~Martel}, 363--381

\bibitem[{{Kormendy} \& {Richstone}(1995)}]{kor95}
{Kormendy}, J. \& {Richstone}, D. 1995, \araa, 33, 581

\bibitem[{{Lacy} {et~al.}(2004)}]{lac04}
{Lacy}, M. {et~al.} 2004, \apjs, 154, 166

\bibitem[{{Magorrian} {et~al.}(1998)}]{mag98}
{Magorrian}, J. {et~al.} 1998, \aj, 115, 2285

\bibitem[{{Marconi} \& {Hunt}(2003)}]{mar03}
{Marconi}, A. \& {Hunt}, L.~K. 2003, \apjl, 589, L21

\bibitem[{Matthews {et~al.}(1964)Matthews, Morgan, \& Schmidt}]{Matthews:64}
Matthews, T.~A., Morgan, W.~W., \& Schmidt, M. 1964, \apj, 140, 35

\bibitem[{Oke {et~al.}(1995)Oke, Cohen, Carr, Cromer, Dingizian, Harris,
  Labrecque, Lucinio, Schaal, Epps, \& Miller}]{Oke:95}
Oke, J.~B., Cohen, J.~G., Carr, M., Cromer, J., Dingizian, A., Harris, F.~H.,
  Labrecque, S., Lucinio, R., Schaal, W., Epps, H., \& Miller, J. 1995, \pasp,
  107, 375

\bibitem[{{Peng} {et~al.}(2002){Peng}, {Ho}, {Impey}, \& {Rix}}]{pen02}
{Peng}, C.~Y., {Ho}, L.~C., {Impey}, C.~D., \& {Rix}, H. 2002, \aj, 124, 266

\bibitem[{Pentericci {et~al.}(2001)Pentericci, McCarthy, R{\"o}ttgering, Miley,
  {van~Breugel}, \& Fowbury}]{Pentericci:01}
Pentericci, L., McCarthy, P., R{\"o}ttgering, H.~J.~A., Miley, G.~K.,
  {van~Breugel}, W.~J.~M., \& Fowbury, R. 2001, 135, 63

\bibitem[{Pierce {et~al.}(2010)}]{Pierce:10}
Pierce, C.~M. {et~al.} 2010, \mnras, iun press(astro-ph/1002.2365)

\bibitem[{{Rieke} {et~al.}(2004)}]{rie04}
{Rieke}, G.~H. {et~al.} 2004, \apjs, 154, 25

\bibitem[{{Sanders} {et~al.}(2007)}]{san07}
{Sanders}, D.~B. {et~al.} 2007, \apjs, 172, 86

\bibitem[{{Schechter} {et~al.}(1997)}]{Schechter:97}
{Schechter}, P.~L. {et~al.} 1997, \apjl, 475, L85+

\bibitem[{{Schinnerer} {et~al.}(2007)}]{schin07}
{Schinnerer}, E. {et~al.} 2007, \apjs, 172, 46

\bibitem[{{Schneider} {et~al.}(2005)}]{sch05}
{Schneider}, D.~P. {et~al.} 2005, \aj, 130, 367

\bibitem[{{Scoville} {et~al.}(2007)}]{scoville07}
{Scoville}, N. {et~al.} 2007, \apjs, 172, 38

\bibitem[{{S\'ersic}(1968)}]{ser1968}
{S\'ersic}, J.~L. 1968, {Atlas de galaxias australes}, ed. J.~L. S\'ersic

\bibitem[{{Seymour} {et~al.}(2008)}]{sey08}
{Seymour}, N. {et~al.} 2008, \mnras, 386, 1695

\bibitem[{{Stern} {et~al.}(2000)}]{ste00}
{Stern}, D. {et~al.} 2000, \aj, 132, 1526

\bibitem[{{Stern} {et~al.}(2005)}]{ste05}
---. 2005, \apj, 631, 163

\bibitem[{{Xu} \& {Komossa}(2009)}]{xu09}
{Xu}, D. \& {Komossa}, S. 2009, \apj, 705, 20

\bibitem[{{York} {et~al.}(2000)}]{yor00}
{York}, D.~G. {et~al.} 2000, \aj, 120, 1579

\bibitem[{{Young} {et~al.}(2008){Young}, {Elvis}, \& {Risaliti}}]{you08}
{Young}, M., {Elvis}, M., \& {Risaliti}, G. 2008, \apj, 688, 128

\end{thebibliography}
% \bibliography{mybib}

\end{document}